\documentclass[12pt]{iopart}
\usepackage{iopams}  
\usepackage{comment}
\usepackage{graphicx}
\usepackage{amsmath,amssymb}
\usepackage{booktabs}
\usepackage{subfigure}
\usepackage[flushleft]{threeparttable}

\begin{document}

\title[Efficient training sets for surrogate models of tokamak turbulence]{Efficient training sets for surrogate models of tokamak turbulence with Active Deep Ensembles}
\author{  L. Zanisi$^1$, A. Ho$^{2,3}$, J. Barr$^4$, T. Madula$^4$, J. Citrin$^{2,3}$, S. Pamela$^{1}$, J. Buchanan$^1$, F. Casson$^{1}$, V. Gopakumar$^{1}$ and JET contributors$^5$ }

\address{
$^1$ United Kingdom Atomic Energy Authority, Culham Centre for Fusion Energy, Abingdon, UK}
\address{$^2$ DIFFER - Dutch Institute for Fundamental Energy Research, Eindhoven, Netherlands}
\address{$^3$Science and Technology of Nuclear Fusion Group, Eindhoven University of Technology, Eindhoven, Netherlands}
\address{$^4$ UCL's Centre for Doctoral Training in Data Intensive Sciences, University College London}
\address{$^5$ See the author list of “Overview of T and D-T results in JET with ITER-like wall” by CF Maggi et al. to be published in Nuclear Fusion Special Issue: Overview and Summary Papers from the 29th Fusion Energy Conference (London, UK, 16-21 October 2023)}


\ead{lorenzo.zanisi@ukaea.uk}
\vspace{10pt}

\begin{abstract}

Model-based plasma scenario development lies at the heart of the design and operation of future fusion powerplants. Including turbulent transport in integrated models is essential for delivering a successful roadmap towards operation of ITER and the design of DEMO-class devices. Given the highly iterative nature of integrated models, fast machine-learning-based surrogates of turbulent transport are fundamental to fulfil the pressing need for faster simulations opening up pulse design, optimization, and flight simulator applications. A significant bottleneck is the generation of suitably large training datasets covering a large volume in parameter space, which can be prohibitively expensive to obtain for higher fidelity codes.

In this work, we propose ADEPT (Active Deep Ensembles for Plasma Turbulence), a physics-informed, two-stage Active Learning strategy to ease this challenge. Active Learning queries a given model by means of an acquisition function that identifies regions where additional data would improve the  surrogate model. We provide a benchmark study using available data from the literature for the QuaLiKiz quasilinear transport model. We demonstrate quantitatively that the physics-informed nature of the proposed workflow reduces the need to perform simulations in stable regions of the parameter space, resulting in significantly improved data efficiency compared to non-physics informed approaches which consider a regression problem over the whole domain. We show an up to a factor of 20 reduction in training dataset size needed to achieve the same performance as random sampling.  We then validate the surrogates on multichannel integrated modelling of ITG-dominated JET scenarios and demonstrate that they recover the performance of QuaLiKiz to better than 10\%. This matches the performance obtained in previous work, but with two orders of magnitude fewer training data points. 

\end{abstract}

%
%
%
%
%

\section{Introduction}
\label{sec:Introduction}
    
    Turbulent transport is the dominant transport mechanism in tokamak plasmas. Understanding and predicting it is essential to achieving fusion power~\cite{aTurbulentTransport-Callen,aToroidalConfinement-Hinton}. 
    Transport models constitute a fundamental tool towards the delivery of ITER, DEMO-class reactors and beyond. However, the computational cost associated to integrating transport models in highly iterative applications such as multichannel integrated models (e.g., JINTRAC, \cite{aJINTRAC-Romanelli}, RAPTOR \cite{Felici2011Raptor}) requires the delivery of fast and accurate surrogates, particularly for many-query applications such as simulation uncertainty quantification, scenario optimization and controller design. Feed-forward neural network (NN) surrogate models of the quasi-linear gyrokinetic model QuaLiKiz \cite{Bourdelle2015,Citrin2017, plassche2020} and the gyrofluid model TGLF \cite{Staebler2007,Staebler2010, Meneghini2017}, have shown a factor 10$^4$ prediction speedup thus enabling real-time capable profile prediction \cite{Felici2018}, discharge optimisation studies \cite{VanMulders2021} and integrated core-pedestal transport models \cite{Meneghini2017,Meneghini2020} at a fraction of the computational cost.

     Due to the cost associated to retrieving large simulation databases to be adopted as training sets for these neural networks, previous works have focused on spanning a small volume in the input space. This restricts some of the current applications to small dimensionality and narrow range in parameter space \cite{RodriguezFernandez2022,Farca2022}, or medium dimensionality \cite{plassche2020}, also sometimes based on experiments \cite{Meneghini2017,jetexp}. 
    At the same time, in \cite{Narita2021}, where linear GKW~\cite{aGKW-Peeters2009} simulations were used to derive semi-empirical saturation rules based on JT60 discharges, an increase in data availability was indicated as a major contributor to the success of the derived reduced-order model in integrated models. In recent work on developing bespoke gyrokinetic surrogates for ITER \cite{Citrin_2023}, increased data efficiency was also identified as a top priority for the development of surrogate models of higher fidelity gyrokinetic codes.

    Big data is not always necessarily informative. Indeed, current datasets obtained from experimental parameter spaces are severely oversampled. For example, \cite{aTwoStep-Kremers} devised a clustering algorithm for the dataset presented in \cite{jetexp}, demonstrating that a performing surrogate can be trained on a carefully selected subsample of the full dataset, with up to a factor of 10 reduction in training set size. Thus, the amount of information needed to obtain an actionable surrogate is contained in a significantly smaller subset of current gyrokinetic databases. While extremely useful to uncover the oversampling problem typical of current approaches, the work by \cite{aTwoStep-Kremers} was performed \emph{a-posteriori}, once the costly training set had already been generated. 
    
    Moreover, by the nature of the critical threshold characteristic of tokamak turbulence, not all plasma states result in unstable modes. In previous work \cite{plassche2020,jetexp}, the consistency of the surrogate with the critical threshold behaviour was enforced by means of a physics-based loss function that encouraged a controlled extrapolation to negative values where the true output fluxes were null. Negative predictions would then be clipped to zero at inference time. Although effective, this strategy lacked in efficiency as it resulted in a large fraction of the computational budget being spent to obtain stable modes (roughly 40\% in \cite{jetexp} across all the electrostatic modes resolved). Instead, we hypothesise that the boundary manifold between stable and unstable inputs may be learned more efficiently using a separate surrogate model. This idea first appeared in our previous work \cite{barr2022}, and it was developed concurrently by Hornsby et al. \cite{hornsby2023gaussian} for data-efficient surrogates of micro-tearing modes.
     
    This study proposes to build NN surrogate models of gyrokinetic turbulence by leveraging Active Learning (AL, \cite{Aggarwal_survey}) methods. Active Learning is a sequential sampling strategy that queries an expensive black box function (in our case a gyrokinetic model) by means of an acquisition function that identifies regions where additional data would improve the NN performance. Contrary to Bayesian Optimisation approaches, which aim to perform sequential optimisation with only a few function evaluations (see for example \cite{RodriguezFernandez2022,aBO-Jarvinen2022,aBayesianPlasmaBoundary-Skvara,chung2020offline} for applications relevant to Fusion), Active Learning enables learning of the function to be approximated over the entire parameter space.
    
    Here we develop ADEPT (Active Deep Ensembles for Plasma Turbulence), a two-stage AL framework where a surrogate of the critical gradient threshold in the form of a classifier determines whether a given input will result in growing modes, and a regressor predicts the output turbulent transport fluxes. We focus on an acquisition function that queries inputs for which the output uncertainty of the NN is highest, thus maximising informativeness \cite{MacKay1992}.  Deep Ensembles \cite{lakshminarayanan2017simple}, which provide state-of-the-art uncertainty quantification capabilities for NNs, are adopted as the surrogate model.

    We provide a demonstration of the ADEPT pipeline using an existing large database of QuaLiKiz simulations obtained from JET inputs \cite{jetexp}. For this proof-of-concept work we focus on ITG turbulence only. As the input-output mappings are already available in the dataset, we can easily test the performance of the two-stage workflow. Explicit integration of gyrokinetic models in ADEPT will follow in upcoming work.
    
    The paper outline is as follows. We describe the dataset in Section \ref{sec:data}, we introduce the ADEPT methodology in Section \ref{subsec:AL} and we outline the integrated modelling framework in Section \ref{sec:IntegratedModelling}. In Section \ref{subsec:ReducedTrainingSets} we give the first main result of the paper. We demonstrate that even only the inclusion of the classifier stage and the adoption of more powerful deep learning models such as Deep Ensembles results in actionable performance for turbulent transport surrogates with around 200,000 simulations for 15 input dimensions, that is a two order of magnitude reduction from the original dataset. The physics-informed nature of the proposed sampling strategy only queries a minority of the inputs in the stable regions, which are instead dominant in the original dataset, thus enabling the surrogate to focus on accurate modelling of non-zero transport fluxes. Sequentially building the training dataset via AL results in a further large reduction in training sample size. In Section \ref{sec:validation} we validate  ADEPT on a representative parameter scan and on integrated modelling of ITG-dominated JET scenarios. We find that ADEPT and previous work \cite{jetexp} agree with JINTRAC runs that adopt the original QuaLiKiz model to better than 10\%, albeit ADEPT was trained with two orders of magnitude less data compared to the surrogates in \cite{jetexp}. Finally, in Section \ref{sec:Conclusion} we discuss the results obtained, identify remaining issues and propose potential solutions to be explored in future work.
    

\section{Data}
\label{sec:data}
We use the existing JET-Exp-15D dataset devised in \cite{jetexp,Ho2019}. The dataset contains the input-output mappings of the QuaLiKiz \cite{Bourdelle2015,Citrin2017} quasilinear model. The inputs are based on 2135 JET experimental discharges including a variety of plasma scenarios, augmented taking into account measurement uncertainties for the parameters that turbulence is most sensitive to. The dataset generation took approximately 150kCPUh.

The input space is 15-dimensional and it includes: the species charge number, the species mass number, the fractional species density, the logarithmic electron density gradient, the ion and electron temperature gradients, the rotation Mach number, the rotation gradient, the radial coordinate, the tokamak aspect ratio, 
 the safety factor, the magnetic shear, the pressure gradient (via $\alpha_{MHD}$) the collisionality and the ExB shearing rate. The output encompasses the multiple channels of transport of ITG, ETG and TEM turbulence obtained from QuaLiKiz. The raw dataset produced from all the available inputs was subjected to consistency checks to either enforce physical consistency within the data (i.e., ambipolar particle fluxes, consistency between the predicted fluxes and the fluxes calculated from combining diffusive and convective terms computed separately) or discard abnormally large heat fluxes and abnormally small particle fluxes; see Table 6 of \cite{jetexp} for more details.

In the remainder of this work the focus will be on ITG turbulence, for which only less than 25\% of inputs in the JET-Exp-15D dataset develop turbulent transport.  The transport fluxes considered (in GyroBohm units, cfr. Table 3 of \cite{jetexp}) are the heat flux of ions ($q_{i,ITG}$) the heat flux of electrons  ($q_{e,ITG}$) the momentum flux of ions ($\Pi_{i,ITG}$) the particle flux of electrons ($\Gamma_{e,ITG}$) and the particle flux of ions ($\Gamma_{i,ITG}$).


\section{Data-efficient surrogate models}
\label{sec:SurrogateModels}
\subsection{Active Learning}
\label{subsec:AL}

\subsubsection{Basics}
\label{subsubsec:ALBasics}

Active Learning (AL, e.g., \cite{Aggarwal_survey} for a review) is a sampling strategy that aims at reducing the amount of training data needed to obtain a performing surrogate. An AL system comprises three components: a learner, an oracle and a query strategy. The learner is a ML method, such as a NN or Gaussian Process \cite{Rasmussen2004}, that improves its performance as more data is collected from the oracle according to the query strategy. The decision on which learner to use depends on the nature of the problem: Gaussian Processes are more suitable in the low-data regime, while NNs are more effective in the big-data limit.  The oracle is a costly data acquisition system that provides the training data for the learner; the oracle might be, for example, a simulator (which is the case this paper is focused on) or a human annotator (such as in the GalaxyZoo project, \cite{Galaxy_zoo_AL}). The main focus of the AL literature is on defining efficient query strategies \cite{Aggarwal_survey}.

AL can be applied in a pool setting and a streaming setting. In the first case, the query strategy acts on a pre-existing pool of unlabelled data (that is, for which only inputs are available but outputs are unavailable) and the distribution of the input space is fixed, while in the second setting a decision on which data to focus the labelling effort is made on a source of streaming data, potentially from a non-stationary distribution. Although digital twinning applications involving building surrogate models off streaming data from fusion devices may benefit from AL, the aim of this paper is to prove the simpler pool setting. 

\subsubsection{Maximum informativeness and uncertainty sampling}
\label{subsubsec:UncertaintySampling}

The goal of AL is to obtain a machine learning predictive model by identifying training points that are more efficient than random selection. Space-filling methods, such as Latin Hypercube Sampling (LHS,\cite{McKay1979}), have been shown to improve upon random selection, however space-filling algorithms sample the input space just once, and therefore do not account for potential redundancy in the information provided by different inputs. A more efficient query strategy consists in maximising the informativeness of the training sample as a whole. The simultaneous placement of N points to obtain optimal coverage of the parameter space of interest is, unfortunately, computationally intractable \cite{Maxentropy_NP_Hard}. Popular alternatives, which include sequential acquisition strategies that account for changes in the model induced by the newly collected training data, are still more advantageous than the fixed design space offered by space-filling algorithms.

The sequential strategy proposed in \cite{MacKay1992} queries the inputs for which the surrogate model's predictive uncertainty is largest,
    \begin{align}
    x_{query} &= \underset{ x \in \mathcal{U} }{arg \ max} \ \sigma^2(x; D_{train,t} ),     \label{eq:5}
 \\
    D_{train,t+1} &= x_{query} \cup D_{train,t} 
    \end{align}
where $\sigma(x; D_{train,t})$ is the output uncertainty of a learner trained on a dataset $D_{train,t}$, $t$ is the current iteration and $\mathcal{U}$ is a pool of inputs (e.g., the plasma states) for which the outputs (e.g., the turbulent fluxes) are not available. As indicated in the expression above, the dataset at the next iteration is enriched with data obtained from the query. The uncertainty is the standard deviation of a regression model.

Here, we adopt Batch Mode AL (e.g., \cite{holzmüller2022framework}), which consists in performing the acquisition for the M inputs that rank highest in the model uncertainty. Batch Mode AL is more suitable for NNs, as retraining a NN with just one new sample is impractical.

The literature on AL strategies is vast (see \cite{Ren_BMDAL_review,Aggarwal_survey} for two excellent reviews). In the following, we will adopt the acquisition function in eq. \ref{eq:5} for the following reasons. First, it is good practice to develop surrogate models that offer uncertainty estimates on their predictions, especially in view of incorporating surrogates of gyrokinetic models, the topic of this paper, into integrated suites to enable uncertainty quantification studies. Moreover, the implementation of uncertainty sampling by exploiting surrogate models with such capabilities is trivial and, as we will show, it performs well in practice. Furthermore, while conceptually very simple, uncertainty-driven AL is widely used with great success in other fields, such as, for example, drug discovery \cite{Soleyimani21}. 

As a final note, it is worth pointing out that AL tends to induce a shift between the distribution of the unlabelled pool $\mathcal{U}$ and the that of the training set over time, as only the most informative points are selected for labelling \cite{MacKay1992,ALBias}. It is therefore crucial to ensure that the NN uncertainties are well-calibrated also out of distribution. A discussion on this matter is carried out in Section \ref{subsec:deepensembles}. 

\subsection{Physics-informed Active Learning for gyrokinetic models with ADEPT}
\label{subsec:ALGK}
\begin{figure}
    \centering
    \includegraphics[width=0.95\textwidth]{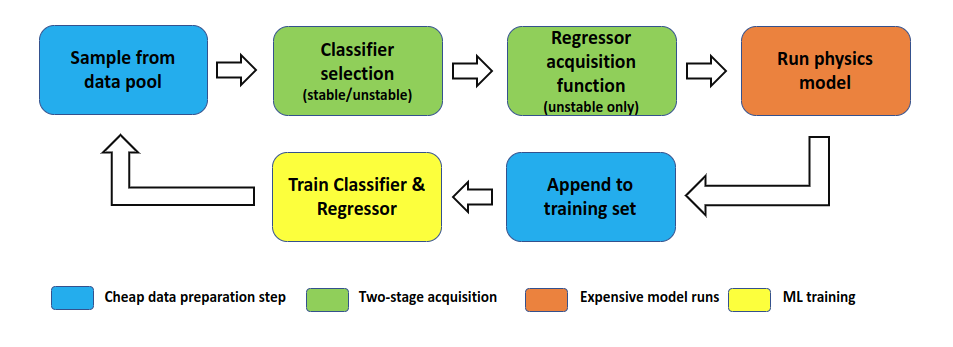}
    \caption{Schematic diagram of the two-stage physics-informed AL workflow used in this work. Given a data pool for which only inputs are available, a classifier evaluates the likelihood of a given input in the pool resulting in unstable modes. The acquisition function is evaluated on the unstable inputs, and a batch of the most uncertain ones are selected to be run through the gyrokinetic model. The newly obtained input-output mappings are used to train both NNs. This strategy is repeated until the computational budget has been exhausted or the performance of the surrogates is deemed actionable. }
    \label{fig:1}
\end{figure}
Linear gyrokinetic turbulence exhibits a critical gradient behaviour, whereby growing modes and the resulting turbulent transport are triggered only above a certain threshold in the driving gradients that depends on the plasma conditions. This creates a further complication for surrogate models, as the fluxes predicted need to be exactly zero in the stable region to avoid the presence of spurious transport that would alter the predictions of integrated models. In previous work \cite{plassche2020} showed that the sharp transition between the stable and unstable region is smoothed out in naive approaches where a single regressor surrogate model is trained on the entire space. The solution proposed in \cite{plassche2020} was to identify the critical gradient threshold was by encouraging a NN to predict negative values whenever the true flux was null. These were then clipped to zero for use in the integrated model. For positive fluxes, instead, the NN would be trained using a standard Mean Squared Error loss function. \cite{plassche2020} showed that a NN surrogate that does not account for the critical gradient behaviour of gyrokinetic turbulence leads to oversmoothing around the critical gradient and therefore overpredicts transport. The physics-informed training adopted in \cite{plassche2020, jetexp} elegantly enables the NN to perform both classification tasks (i.e. whether an input results in growing modes) and regression tasks (to predict the turbulent fluxes). While effective, \cite{plassche2020}'s method results in the computational budget spent to obtain the training set to be overly focused on points well within the stable region of the input space.

Below we propose ADEPT (Active Deep Ensembles for Plasma Turbulence)\footnote{We adopt the \texttt{PyTorch} library.,} a two-stage, physics-informed active learning strategy that delivers a significant reduction in the volume of data required to train performing surrogates. Contrary to previous work, we assign the classification and regression tasks to two separate neural networks. This setup preserves the physics-informed nature of the framework proposed by \cite{plassche2020}, but it splits the burden of identifying the critical gradients and regressing to turbulent fluxes between two highly specialised NNs. Given a data pool $\mathcal{U}$ of inputs, a NN classifier and a NN regressor are pretrained on a small (20,000 points) random sample of data for which the input-output mapping is available. This pretraining allows to capture a general initial representation of the space. Hereafter, for each iteration, the networks and the labelled dataset are updated following the strategy shown in Figure \ref{fig:1}:
\begin{itemize}
    \item The classifier is tasked with screening a sample of candidate points  in the data pool $\mathcal{U}$. This is the physics-informed stage of the workflow. The entire pool may be screened, but this may slow down the acquisition process in the case of large data pools, such as that of the JET-Exp-15D dataset. Therefore the classifier is used to evaluate 300,000 inputs randomly sampled from the pool;
    \item The acquisition function in eq. \ref{eq:5} uses the regressor's uncertainty (in our case, the epistemic uncertainty in eq. \ref{eq:10}) to select a batch of size \texttt{acquisition\_batch} candidates. Extending the acquisition strategy to account for the uncertainty of the classifier will be the subject of future work;
    \item The outputs for the input candidates selected are queried from the model of choice (QuaLiKiz in the case of this paper); 
    \item The newly available input-output mappings are appended to the training data;
    \item Both the regressor and the classifier NNs are trained again.
\end{itemize}
The loop above is repeated until the computational budget has been exhausted, or the surrogates have reached the desired performance.

As gyrokinetic turbulence involves multichannel transport, it is necessary to maximise the information gain for all the fluxes, and therefore the acquisition function in eq. \ref{eq:5} becomes
 \begin{equation}
    x_{query} = \underset{ x \in \mathcal{U} }{arg \ max} \ \sum_{k=1}^{N_{fluxes}} \sigma^2_k(x;  D_{train,t} ).  
     \label{eq:5_sum}
 \end{equation}

\subsection{Surrogate uncertainty via Deep Ensembles and its uses within integrated modelling}
\label{subsec:deepensembles}

It has long been established that NN models give overconfident predictions that are factually wrong (e.g., \cite{nguyen2015deep}). Equipping NNs with a notion of uncertainty in their own predictions has since become a mainstream line of research producing a rich literature \cite{gawlikowski2022survey}. The calibration of NN uncertainties are currently debated in the community \cite{pmlr-v70-guo17a}, with new frequentist methods on the rise (\cite{angelopoulos2022gentle} and references therein). In particular, although NN uncertainties generally increase moving away from the training distribution across all uncertainty estimation methods, the issue of their calibration remains a point of concern.  


\cite{lakshminarayanan2017simple} proposed to train NNs using \emph{proper scoring rules} \cite{proper_scoring_rule} to obtain calibrated uncertainties. Given a NN approximation $p_\theta(y|x)$\footnote{Where $\theta$ are the parameters of the NN.} of a distribution that approximates the truth, $q(y|x)$, a scoring rule $\mathcal{S}\bigl(p_\theta, (x,y)\bigr)$ assigns to a learned supervised model a score based on the quality of the model's uncertainty for a particular input-output pair. A scoring rule can be formalised as global metric by integrating over the full probability space, 
\begin{equation}
    S(p_\theta,q) = \int q(x,y)\mathcal{S}\bigl(p_\theta, (x,y)\bigr)dxdy.
    \label{eq:properscoring}
\end{equation}
A scoring rule is \emph{strictly proper} if $S(p_\theta, q)\leq S(q,q)$, that is, the learned approximation $p_\theta$ is best only in the case where it perfectly reproduces $q$, that is when $p_\theta=q$.

A NN trained with a proper scoring rule is encouraged to provide better calibrated uncertainties compared to one trained on MSE. The standard MSE loss routinely used to train NNs (see eq. \ref{eq:6}) is not a strictly proper scoring rule \cite{Brcker2007}, and therefore cannot provide calibrated uncertainties out of the box.  Instead, \cite{lakshminarayanan2017simple} showed that the log-likelihood of the data under a learned NN, $\log{p_\theta(y|x)}$, is always a strictly proper scoring rule and it provides calibrated uncertainties also in practice. Ensembles of NNs trained with a proper scoring rule are termed \emph{Deep Ensembles}. If the ensemble is treated as a uniformly weighted mixture model, then the proper scoring rule for the ensemble is 
\begin{equation}
    NLL = -\frac{1}{M}\sum_{k=1}^M \log{p_{\theta_k}}(y|x)
\end{equation}
where we have taken the negative of the log-likelihood as an objective to minimise.

For classification, the usual binary cross-entropy loss is also a proper scoring rule, and therefore deep ensembles and regular NN ensembles coincide. 
For regression problems, the Gaussian negative log-likelihood below is a proper scoring rule, 
\begin{equation}
    -\log{p_\theta(y|x)} = \frac{\log{\sigma_\theta^2(x)}}{2} + \frac{\bigl(y-\mu_\theta(x) \bigr)^2}{2\sigma_\theta^2(x)} + const.
    \label{eq:loglikelihood}
\end{equation}
A NN with two output neurons trained with the objective above will explicitly learn the mean $\mu_\theta(x)$ and variance $\sigma_\theta^2(x)$, where the suffix indicates that these quantities are parametrised by the same NN with parameters $\theta$. With this expression, the NN is encouraged to learn that, in order to have a low variance to minimise the first term of eq. \ref{eq:loglikelihood}, the predictions $\mu_\theta$ need to be very accurate to keep the second term small.

The mean $\mu_E$ and variance $\sigma_E$ of the deep ensemble as a whole can be computed under the assumption of a uniformly weighted mixture of M members:
\begin{align}
    \mu_E &= \frac{1}{M}\sum_{k=1}^M \mu_{\theta_k} \label{eq:9}\\
    \sigma_{E}^2 &=
    \underbrace{\frac{1}{M}\sum_{k=1}^M \sigma_{\theta_k}^2}_\text{aleatoric}
    + \underbrace{\bigl ( \frac{1}{M}\sum_{k=1}^M \mu_{\theta_k}^2 \bigr ) - \mu_E^2.}_\text{epistemic} \label{eq:10}
\end{align}

On the other hand, \cite{jetexp} used an NN ensembling slightly different approach compared to Deep Ensemble to obtain a notion of uncertainty. The approach consisted in training a committee of ten NNs with identical architecture but different random initialisation, and the mean and variance of the predictions were then used for downstream applications. The NNs in \cite{jetexp} were trained to minimise the Mean Squared Error (MSE) between each NN prediction, $\hat{y}$ and the target, $y_{true}$,
\begin{equation}
    MSE = \frac{1}{N}  \sum_{i=1}^N ( \hat{y}_i-y_{true,i} )^2.
    \label{eq:6}
\end{equation}
Note that the expression in eq. \ref{eq:loglikelihood}  allows for heteroskedasticity in the variance estimate (i.e. the variance can vary based on each individual input, and this is explicitly modelled). It is important to realise that, without this feature, the expressions in eq. \ref{eq:loglikelihood} and eq. \ref{eq:6} would coincide (up to a constant) after identifying $\mu_\theta\equiv \hat{y}$. Although this may seem only a subtle difference between Deep Ensembles and regular NN committees, the objective in eq. \ref{eq:6} does not explicitly capture NN uncertainty. Therefore, the uncertainties obtained by considering the standard deviation of the ensemble outputs are not guaranteed to be valid. Instead, training Deep Ensembles involves the optimisation of the negative log likelihood, which improves MSE with the constraint of fitting sensible uncertainty estimates. Hence, Deep Ensembles strike a balance between uncertainty quantification capabilities and predictive power, which are both equally important in downstream applications.

The variance of a Deep Ensemble regressor (eq. \ref{eq:10}) is composed of two contributions. The first one is the average variance between all members. The second one is the variance of the means of the ensemble, as shown in the last two terms on the right hand side of eq. \ref{eq:10}. 
The uncertainty of the deep ensembles (eq. \ref{eq:10}) is sometimes interpreted as the sum of the epistemic uncertainty (i.e. the uncertainty of the model) and the aleatoric uncertainty (i.e. the irreducible noise in the data), e.g. \cite{gustafsson2020evaluating}. The epistemic uncertainty is the natural choice to use in the acquisition function, as we seek to improve the inherent accuracy of the model regardless of data noise \cite{Yudin_2023}. Conversely, the total uncertainty should be used to assess how much trust should be placed in the surrogate predictions for downstream applications such as integrated models.  

Uncertainty quantification capabilities are also a natural feature of the classifier NN. The confidence of the classifier can be defined as its output probability of a point being unstable. Probabilities close to a value of 0.5 inform downstream applications that performing a run of the original QuaLiKiz model is recommended. Entropy \cite{shannonMathematicalTheoryCommunication1948}, which measures the disagreement between the members of the ensemble, may also be used as an information-theoretical measure of uncertainty:
\begin{equation}
    \mathcal{H}(x) = - \sum_{i=1}^M p_i(x)\log{p_i(x)},
    \label{eq:entropy}
\end{equation}
where $p_i(x)$ is the output probability of the $i-th$ ensemble member. Both probability and entropy will be shown as measures of uncertainty for the classifier for a few parameter scans in Section \ref{subsec:ParameterScans}.




\subsection{Details of the training procedure}
\label{subsec:TrainingDetails}
We borrow from \cite{plassche2020} the idea of fitting NN surrogate models to the ``leading flux`` of a given turbulence type and the flux ratio between the leading flux and the secondary fluxes. This methodology was devised to ensure the same critical gradient behaviour for all fluxes of a given turbulent mode. While this is not strictly necessary in our case, as the classifier takes care of identifying the critical gradients, we opted for this option to minimise changes in the JINTRAC integration.

We train a suite of deep ensembles, each regressing to one turbulent flux, and one deep ensemble classifier for the stability boundary. We adopt 5 ensembles per model, each consisting of 8 layers with 512 parameters each and \texttt{ReLU} activation functions. Each model is trained for 200 epochs with 100 epochs of patience, a weight decay of $\lambda$=$10^{-4}$ and a \texttt{batch\_size} of 512. 
For the regressor we adopt the NLL loss and for the classifier the binary crossentropy loss. Each \texttt{acquisition\_batch} consists of 512 training samples, doubling every 30 acquisitions due to the costs associated with retraining NNs on large datasets. 

Extensive hyperparameter tuning was not performed in this study. Optimizing hyperparameters should ideally occur during each iteration of AL. However, even in AL research this is rarely done due to its high computational cost. Performing hyperparameter tuning at every iteration can be prohibitively time-consuming, especially as the training dataset grows. However, the extra computational cost incurred is expected to be small compared to the data acquisition for expensive high fidelity codes (e.g., \cite{JenkoGENE, aGKW-Peeters2009}).

\section{Integrated modelling}
\label{sec:IntegratedModelling}

 The JINTRAC integrated modelling suite ~\cite{aJINTRAC-Romanelli} was chosen for this study both due its history of integration with QLKNN~\cite{jetexp,plassche2020} and its relevance for ITER scenarios~\cite{aITER-JINTRAC-Asp2022}. 
The JINTRAC integrated model test cases and settings used in this study were taken from those used to validate QLKNN-jetexp~\cite{jetexp}, specifically selecting:
\begin{itemize}
    \item the H-mode carbon wall discharge (JET\#73342) \cite{Citrin2017};
    \item and the H-mode beryllium wall discharge (JET\#92436) \cite{Ho2019}.
\end{itemize}
A summary of the major JINTRAC settings used for these test cases are provided in Table~\ref{tbl:JETTOSettingsSummary}.  In the following Sections we will compare the results of adopting either ADEPT, QLKNN-jetexp and the original QuaLiKiz within JINTRAC. Specifically, the critical gradient threshold in ADEPT will be estimated using the trained classifier, while it is estimated according to the methodology summarised in Section \ref{subsec:ALGK} for the QLKNN-jetexp surrogates.

\begin{table*}[tbp]
	\centering
	\begin{threeparttable}
		\caption{Summary table of most pertinent JINTRAC settings of the base case simulation.}
		\begin{tabular}{l|ccc}
			\toprule[1.5pt]
			& JET\#73342 & JET\#92436  \\
			\midrule
			Description & H-mode & H-mode \\
			Simulation type & Stationary & Stationary \\
			\midrule
			\# of grid points & 51 & 101  \\
			Plasma time & 20.75 -- 22.75~s & 10.0 -- 12.0~s \\
			Sim. boundary ($\rho_{\text{tor}}$) & 0.85 & 0.85 \\
            Equilibrium & Fixed & ESCO  \\
            Neoclassical transport model & NCLASS & NCLASS  \\
            Neutral particle model & None & None  \\
            NBI source model & Fixed & Fixed  \\
            ICRH source model & Fixed & Fixed  \\
			Impurity species & C & Be, Ni, W  \\
            Impurity transport model & None & SANCO  \\
            QuaLiKiz region & 0.15 -- 0.85 & 0.15 -- 0.85  \\
            QuaLiKiz rot. option & 2 & 2  \\ 
            Part. trans. option\tnote{1} & 4 & 4 \\
			Momentum profile & Fixed & Predicted  \\
			\bottomrule
		\end{tabular}
		\begin{tablenotes}
			\item[1] The particle transport options are only applicable when using the QLKNN model. Further details about the different options available within QLKNN are given in Ref.~\cite{jetexp}.
		\end{tablenotes}
		\label{tbl:JETTOSettingsSummary}
	\end{threeparttable}
\end{table*}

As outlined in Section 4.2 of Ref. \cite{jetexp}, the ion transport coefficients for the JET-Exp-15D dataset were derived only for a pure deuterium plasma, and therefore some assumptions need to be made to model impurity transport. While the differences in heat transport among different ion species can usually be neglected, this is not typically true for particle fluxes \cite{Marin2020}.

A first condition to allow treatment of the particle transport coefficients of impurities stems from the ambipolarity constraint,
\begin{equation}
\label{eq:ambipolarity}
    \boldsymbol{\Gamma_e} = \sum_i \boldsymbol{\Gamma_i}Z_i.
\end{equation}

However, a second condition needs to be specified for eq. \ref{eq:ambipolarity} to admit a unique solution. In this work, we follow Ref. \cite{jetexp} and assume a proportionality between the electron and the impurities particle fluxes:
\begin{equation}
    \label{eq:impurities}
 \boldsymbol{\Gamma_i} = \boldsymbol{\Gamma_e}\frac{n_i}{n_e}.    
\end{equation}


ADEPT generates surrogate models that inherently include a measure of uncertainty. This characteristic can be leveraged in integrated modeling to evaluate the level of confidence that should be placed in the surrogate model's predictions, and perform a QuaLiKiz run whenever the uncertainty of surrogate is not considered acceptable. An in-depth study of the impact of the precise acceptance threshold on the integrated modelling results is outside the scope of this work, but it is highly recommended for future investigation. In particular, in this study the average predictions of the surrogates are used regardless of the surrogate uncertainty. 

\section{Results: data-efficient training sets with active learning }
\label{subsec:ReducedTrainingSets}

\begin{figure}
    \centering
    \includegraphics[width=0.8\textwidth]{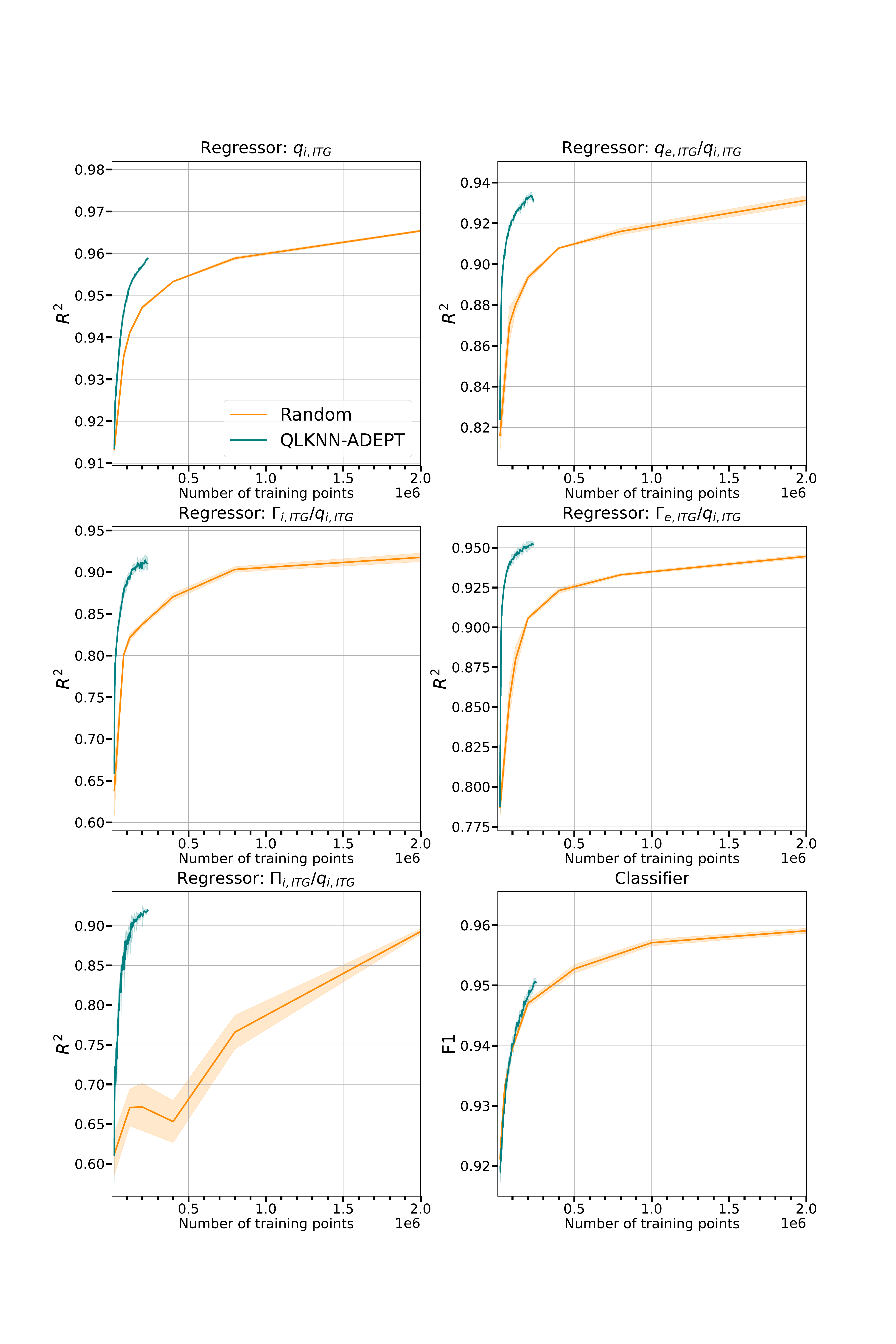}
    \caption{ADEPT vs random sampling performance. We use the $R^2$ for the deep ensembles regressing to the electron heat flux and the flux ratios involving the ion heat flux and the ion momentum flux. The $F1$ score for the stability boundary classifier is also shown. The shaded areas represent the standard deviation of 5 runs with different random seed. Active Learning improves data efficiency by at least factor of 2 (and up to a factor of 20) compared random sampling. ADEPT acquisitions were run until exhaustion of the computational budget (36 hours).}
    \label{fig:AL_results}
\end{figure}

An important benchmark of performance of any Active Learning strategy is given by random sampling. Specifically, the extra costs incurred in Active Learning due to retraining the surrogate at every acquisition are justified solely if Active Learning outperforms random sampling, which requires training only once. The metrics used to assess the surrogate performance are described in detail in \ref{app:perfmetrics}.

Figure \ref{fig:AL_results} shows the performance of ADEPT on ITG turbulence compared to random selection as a function of number of training samples collected.  For both ADEPT and random selection the same NN architectures and training hyperparameters were adopted. It can be seen that ADEPT provides up a factor of 20 data reduction compared to random sampling. As shown in Section \ref{subsec:RemovePhysicsInformed}, an important contribution to this success is the inclusion of the classifier stage, which allows for a more data-efficient learning of the manifold where unstable turbulent fluxes develop.

A performance comparison on a test set between ADEPT and the surrogates presented in \cite{jetexp}, which were trained using approximately 20,000,000 data points, is given in Tables \ref{tab:performance_table} and \ref{tab:performance_table_cls}. Although the surrogates in Ref. \cite{jetexp} do not explicitly employ a separate classifier NN to model the critical gradient, they achieve a comparable effect by zeroing out all fluxes when the leading flux is predicted to be negative. Therefore, we can evaluate the F1 performance of these surrogates, as they effectively exhibit classifier-like behavior in this context. It can be seen that the performance of ADEPT in terms of the F1 score is comparable or superior to that of the NNs in Ref. \cite{jetexp}, albeit with two orders of magnitude fewer data. In particular, a high classifier performance in the case of ADEPT is crucial in ensuring that sampling does not occur deep in the stable regions. As a demonstration, we have computed that the contribution of stable inputs to the training set of the classifier is 75\% in the random sampling case (and, indeed, the case of Ref. \cite{jetexp}), while this drops to around 20\% in the case of ADEPT. The surrogates of \cite{jetexp}, instead, feature a poor Precision, showing a high number of non-zero flux predictions in the stable region. The latter surrogates, however, achieve a better Recall, albeit with two orders of magnitude more training data points. ADEPT would need more data to reach the same kind of performance. As further discussed in Section \ref{sec:Conclusion}, the classifier is not currently included in the acquisition function explicitly, which instead will be crucial to improve its data efficiency compared to random sampling. This feature will be explored in future work.

\begin{table}
    \centering
    \begin{tabular}{c|c|c|c|c|c}
     & $q_{i,ITG}$ & $q_{e,ITG}$/$q_{i,ITG}$ & $\Gamma_{i,ITG}$/$q_{i,ITG}$ & $\Gamma_{e,ITG}$/$q_{i,ITG}$ & $\Pi_{i,ITG}$/$q_{i,ITG}$ \\
    \hline \hline
    ADEPT      & 0.9585 & 0.9366 & 0.9174 & 0.9554 & 0.9108 \\
    \hline
    \cite{jetexp} & 0.9518 & 0.9505 & 0.6987 & 0.5268 & 0.9140 \\
    \end{tabular}
    \caption{The performance of ADEPT trained with up to 200,000 samples compared to the NNs presented in \cite{jetexp} (where 20,000,000 samples were used), in terms of $R^2$ score for the fluxes (see \ref{app:perfmetrics} for a description of the performance metrics).}
    \label{tab:performance_table}
\end{table}

\begin{table}
    \centering
    \begin{tabular}{c|c|c|c}
     & F1 & Recall & Precision \\
    \hline \hline
    ADEPT & 0.9504 & 0.9412 & 0.9602 \\
    \hline
    \cite{jetexp} & 0.7791 & 0.9880 & 0.6431 \\
    \end{tabular}
    \caption{The performance of ADEPT trained with up to 200,000 samples compared to the NNs presented in \cite{jetexp} (where 20,000,000 samples were used), for the classifier (see \ref{app:perfmetrics} for a description of the performance metrics).}
    \label{tab:performance_table_cls}
\end{table}

\subsection{The effect of abandoning the physics-informed approach}
\label{subsec:RemovePhysicsInformed}
\begin{figure}
    \centering
    \includegraphics[width=0.9\textwidth]{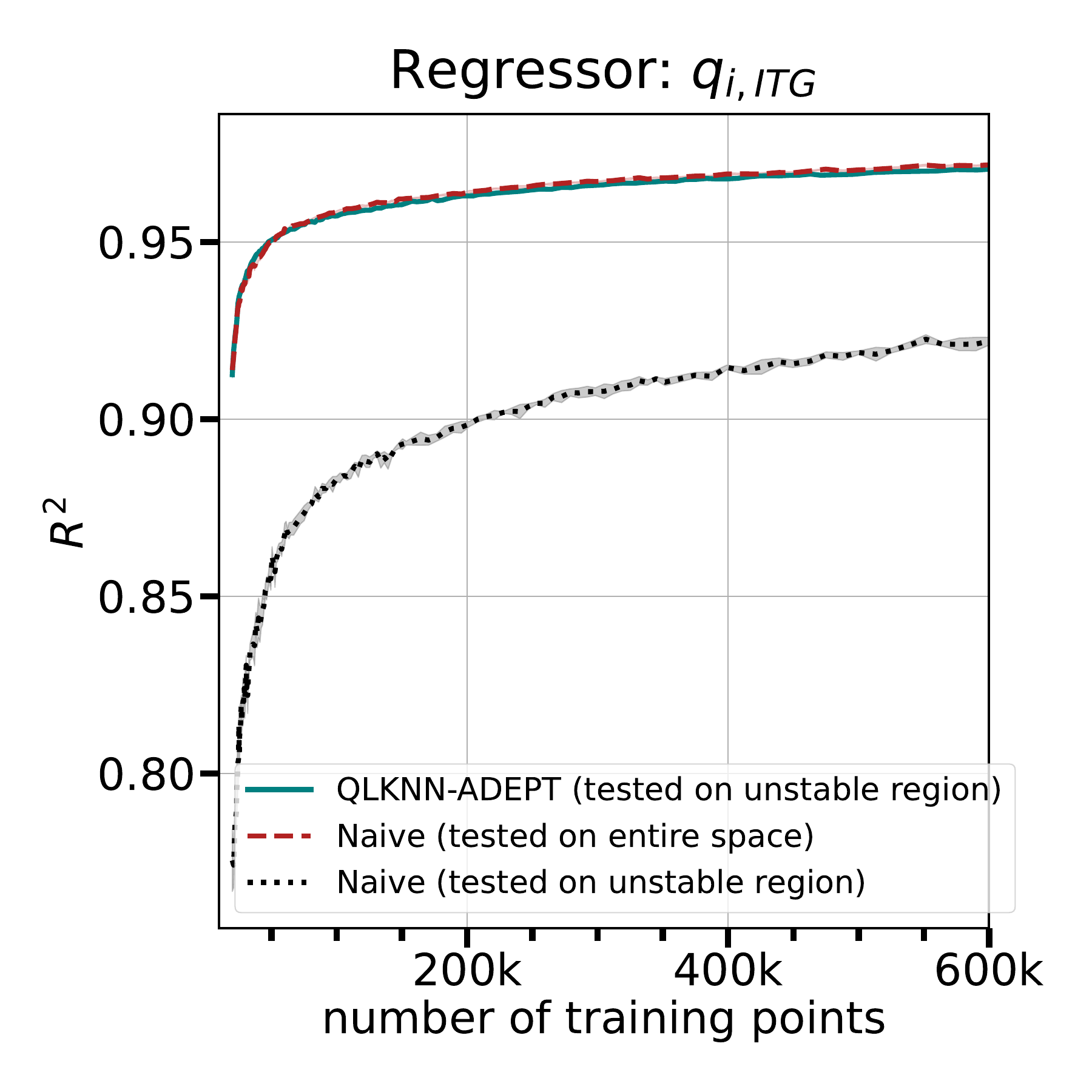}
    \caption{The performance of the regressor in the physics-informed two-stage ADEPT workflow (teal) versus a naive approach where both zero and non-zero fluxes are fit by the same regressor NN (red). It can be seen that, for a fixed amount of training data, when tested on the unstable region only (gray line), the naive approach achieves a much poorer performance than the physics-informed approach.}
    \label{fig:noclassifier}
\end{figure}

The importance of ensuring that the critical gradient of turbulent transport is preserved by surrogates was discussed already in \cite{plassche2020}. The behaviour of one ``naive" regressor surrogate model that predicted all output fluxes, including in the stable region, and without the clipping strategy for negative leading fluxes proposed in \cite{plassche2020}, was shown to oversmooth the critical gradient behaviour and produced unphysical results within integrated modelling.

In this Section we further demonstrate the two following points: (i) providing an estimate for the critical gradient (i.e., utilising a physics-informed approach) results in increased data efficiency within ADEPT compared to naive surrogates and (ii) as a consequence of (i) the seemingly good integrated performance of the naive approach actually results in poor performance in the unstable region compared to ADEPT.

To this end, we performed an experiment where Active Learning was run in a naive fashion, where one regressor was trained on both stable and unstable regions, using only the regressor uncertainty to drive the acquisition. No classifier was used for this experiment. The test set that is natural to use for this method is drawn from the entire space (red line in Figure \ref{fig:noclassifier}) and, at face value, the performance of the naive methodology seems actionable. It is however instructive to inspect the performance solely on the unstable region. Figure \ref{fig:noclassifier} demonstrates that the data efficiency of the naive method degrades significantly when specifically tested on unstable inputs. A crucial observation that justifies the observed behaviour is that the JET-Exp-15D dataset used in this work contains a significant proportion of stable inputs, accounting for over 75\% of the data available for ITG turbulence. Thus, the representation learned by the naive approach is not capable of accurately capturing the mapping for both stable and unstable regions.

On the contrary, the classifier stage of ADEPT helps prevent querying points inside the stable region and instead allows the regressor to focus on the unstable region, thus resulting in improved data efficiency. In line with \cite{plassche2020}, our findings show that integrated performance metrics must be handled with care when informing suitability for downstream applications.

\subsection{Training dynamics dependence on number of fluxes.}
\label{subsec:TrainingDynamics}
\begin{figure}
    \centering
    \includegraphics[width=0.9\textwidth]{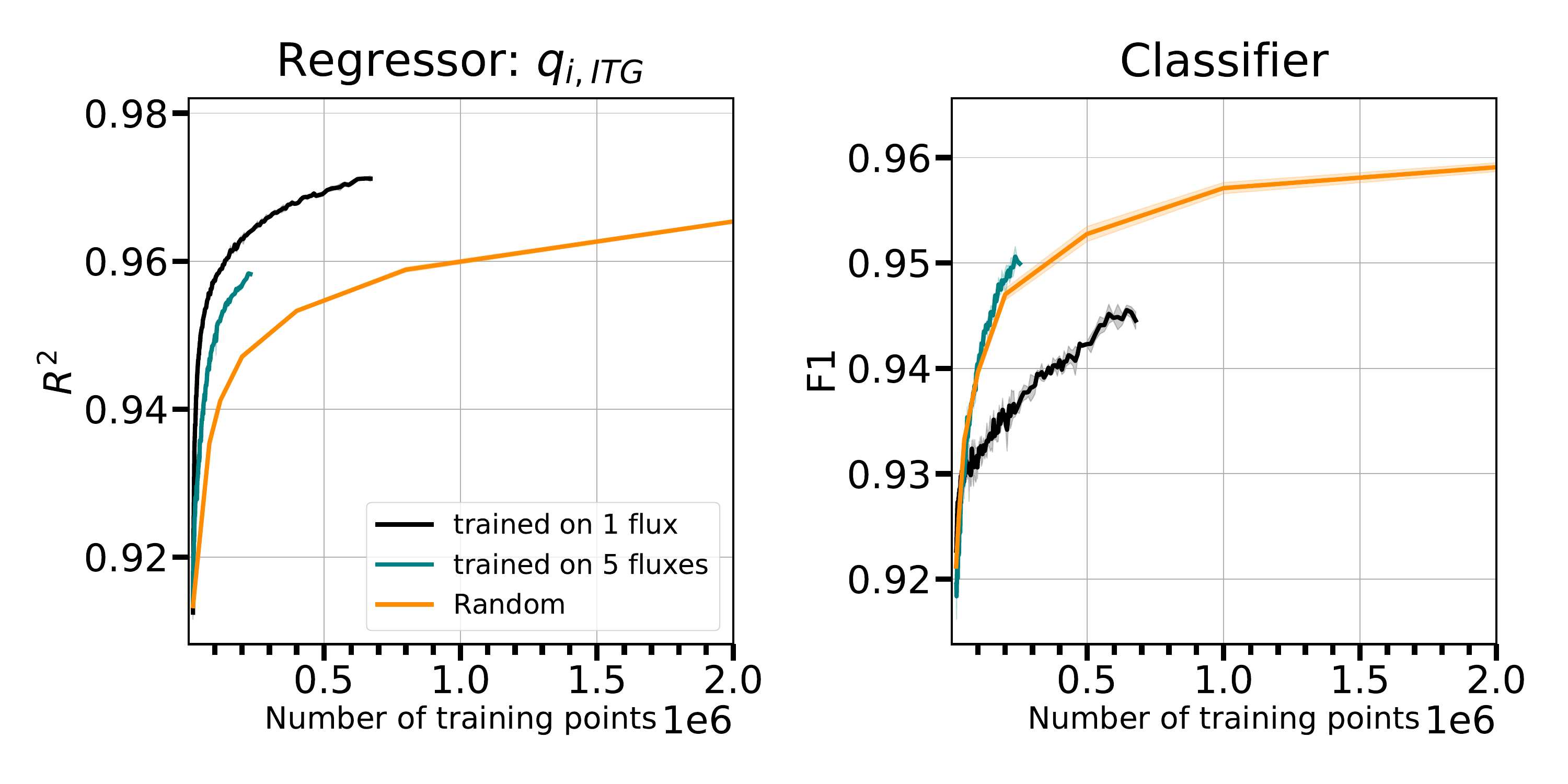}
    \caption{The performance of the ADEPT pipeline for $q_{i,ITG}$ when the surrogates are trained either $q_{i,ITG}$ only (black lines) or on all the five fluxes considered (teal lines, reproduced again from Figure \ref{fig:AL_results} for convenience). The behaviour of both the regressor and classifier depend strongly on the number of fluxes used. ADEPT acquisitions for both the black and teal lines were run until exhaustion of the computational budget (36 hours).}
    \label{fig:Different_num_fluxes}
\end{figure}
The acquisition function in eq. \ref{eq:5_sum} fully accounts for the multichannel nature of gyrokinetic turbulent transport. It is instructive to inspect the training dynamics induced by the training samples collected iteratively by the acquisition function when using a different number of fluxes. 

In Figure \ref{fig:Different_num_fluxes} we show the test performance of the two-stage ADEPT pipeline when only the leading flux, $q_{i,ITG}$, is used. We compare the results to the case where all five fluxes are considered. While the performance on predicting the fluxes in the unstable region is greatly improved compared to the multichannel case, it can be seen that the classifier performance degrades significantly, performing even worse than random sampling. A possible explanation for this behaviour is that in the multichannel case the contribution of the uncertainties from the different fluxes conspire to query a batch that carries high information for the classifier, but not for the regressor surrogate of $q_{i,ITG}$.  

Ultimately, the patterns evident in Figure \ref{fig:Different_num_fluxes} are driven by the acquisition function, which relies solely on the uncertainty of the regressors. We believe that this behavior can be controlled by developing an alternative acquisition function that explicitly takes into account classifier uncertainty.

\section{Results: Validation}
\label{sec:validation}
Bearing in mind that a large-scale evaluation study including uncertainty quantification is outside the scope of the present paper, in this Section we validate ADEPT on parameter scans (Section \ref{subsec:ParameterScans}) and JINTRAC modelling of selected JET discharges (Section \ref{subsec:JETValidation}) as introduced in Section \ref{sec:IntegratedModelling}. We use surrogates that were trained on a final dataset of 200,000 input-output pairs collected using the ADEPT strategy and compare their performance to the work of \cite{jetexp} (QLKNN-jetexp in the following), which were trained using approximately 20,000,000 input-output pairs.

\subsection{Parameter scans}
\label{subsec:ParameterScans}
\begin{figure}
    \centering
    \includegraphics[width=0.9\textwidth]{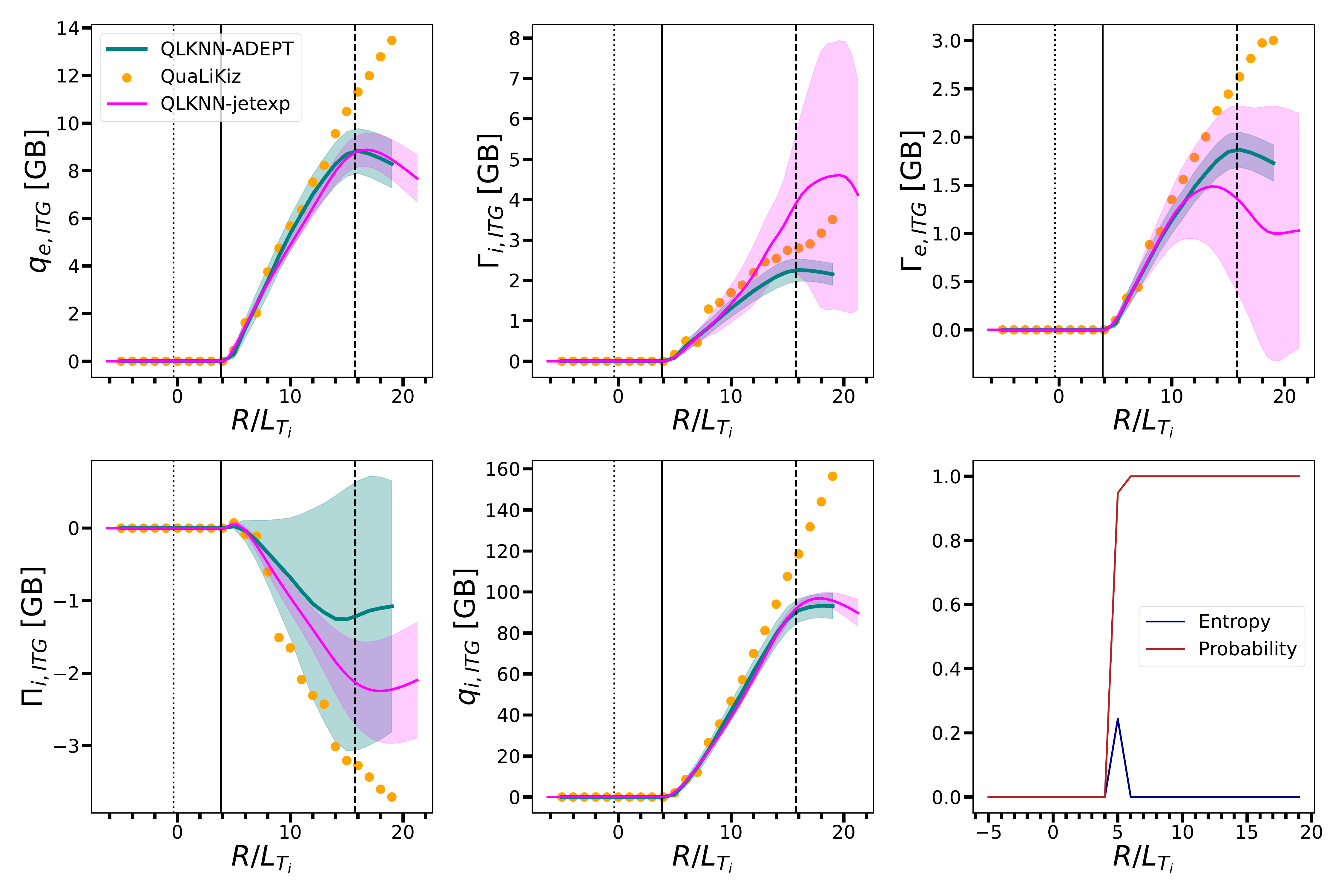}
    \caption{Parameter scans in $R/L_{T_i}$ for the five ITG fluxes used in this work. The QuaLiKiz runs are shown in orange, while the predictions of the surrogates are shown in teal for ADEPT and magenta for QLKNN-jetexp. The shaded areas indicate the 1$\sigma$ confidence levels. Dotted, solid and dashed lines indicate the 2.5\%, 50\% and 97.5\% of the distribution in $R/L_{T_i}$. The bottom right panel shows the uncertainty for the Deep Ensemble classifier in terms of probability of an input being unstable and entropy of the ensemble. Note that the uncertainty estimates provided by the committees in \cite{jetexp} and by ADEPT differ significantly. See main text for discussion.}
    \label{fig:paramscans}
\end{figure}

In this Section, we validate the ADEPT surrogates on parameter scans obtained by running the original QuaLiKiz model. For each output flux, we fix 14 of the 15 input dimensions of the dataset to their median value and we perform a scan in the remaining dimension. Figure \ref{fig:paramscans} shows the parameter scans for $R/L_{T_i}$. Other similar figures can be found in \ref{app:OtherScans}. There is good agreement between the true QuaLiKiz model compared to the NN predictions and their uncertainty on the turbulent fluxes. As expected, the NN uncertainty is larger further away from the bulk of the training distribution for all fluxes. 

Although qualitatively QLKNN-jetexp and ADEPT reproduce the QuaLiKiz trends, there are important differences in how the two approaches perform around the critical gradient. In particular, QLKNN-jetexp tends to provide a smoother behaviour while the two-stage nature of ADEPT results in sometimes too sharp discontinuities (see Figure \ref{fig:otherparamscans_ane}). However, in some instances (see, e.g., Figures \ref{fig:otherparamscans_ate}, \ref{fig:otherparamscans_gammae}) QLKNN-jetexp oversmooths the trends around the critical gradient, albeit it does so out of the training distribution. It is also important to note that the classifier uncertainty for ADEPT peaks around the critical gradient, which is highly desirable as it provides a way to refine the critical gradient estimation - this is a new feature that was not present in QLKNN-jetexp. 

\subsection{Validation on ITG-dominated JET discharges}
\label{subsec:JETValidation}

\begin{figure}
    \centering
    \subfigure[]{
        \includegraphics[width=0.45\textwidth]{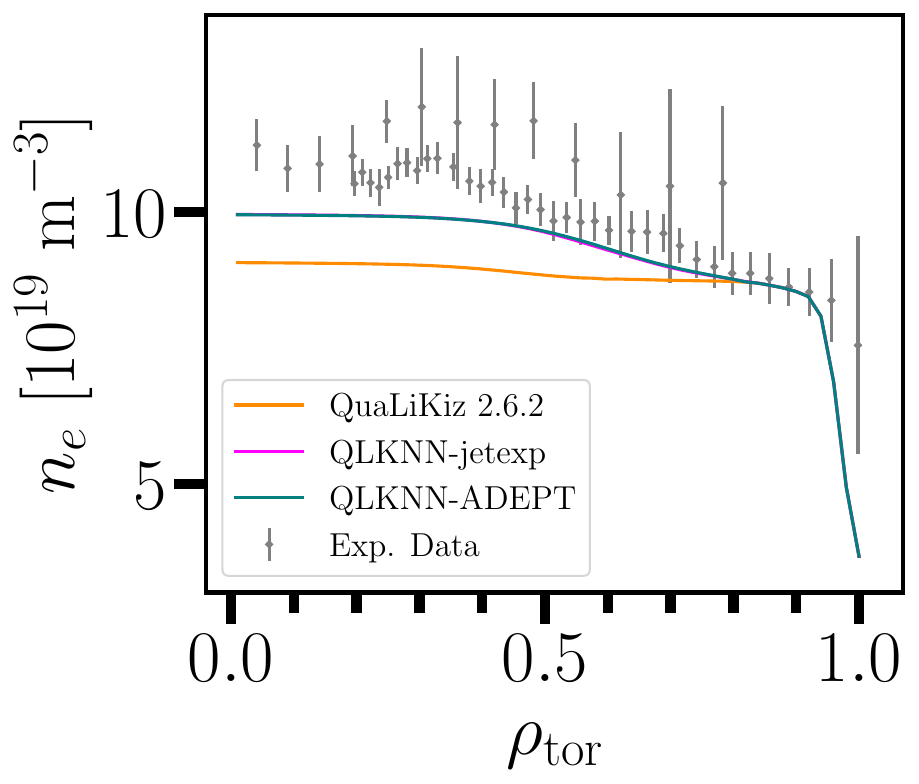}
    }
    \hfill
    \subfigure[]{
        \includegraphics[width=0.45\textwidth]{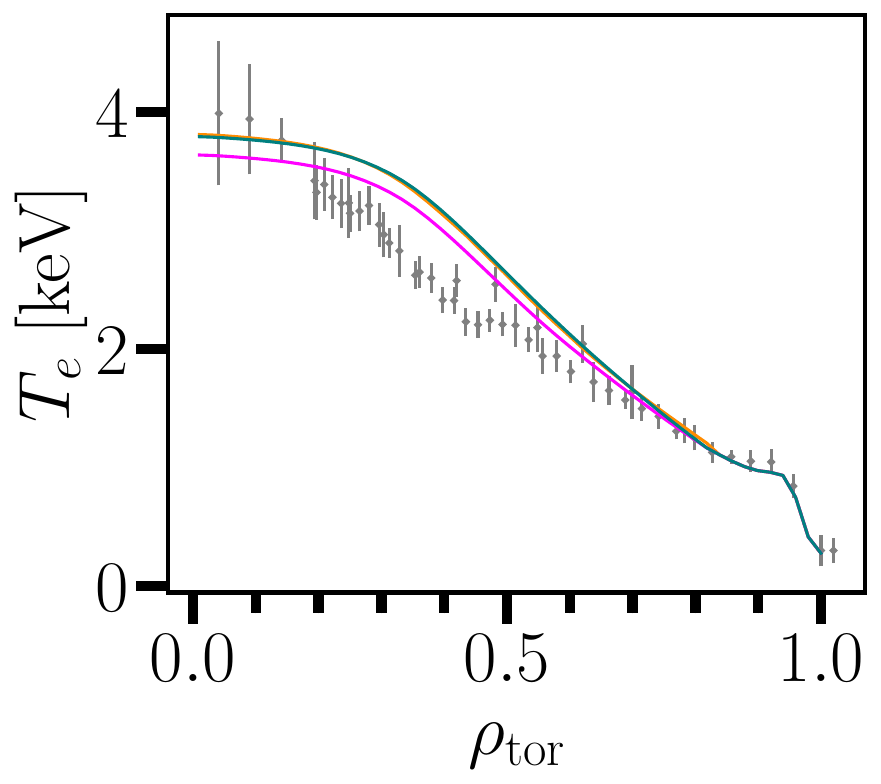}
        
    }
    \vspace{0.5cm} 
    \subfigure[]{
        \includegraphics[width=0.45\textwidth]{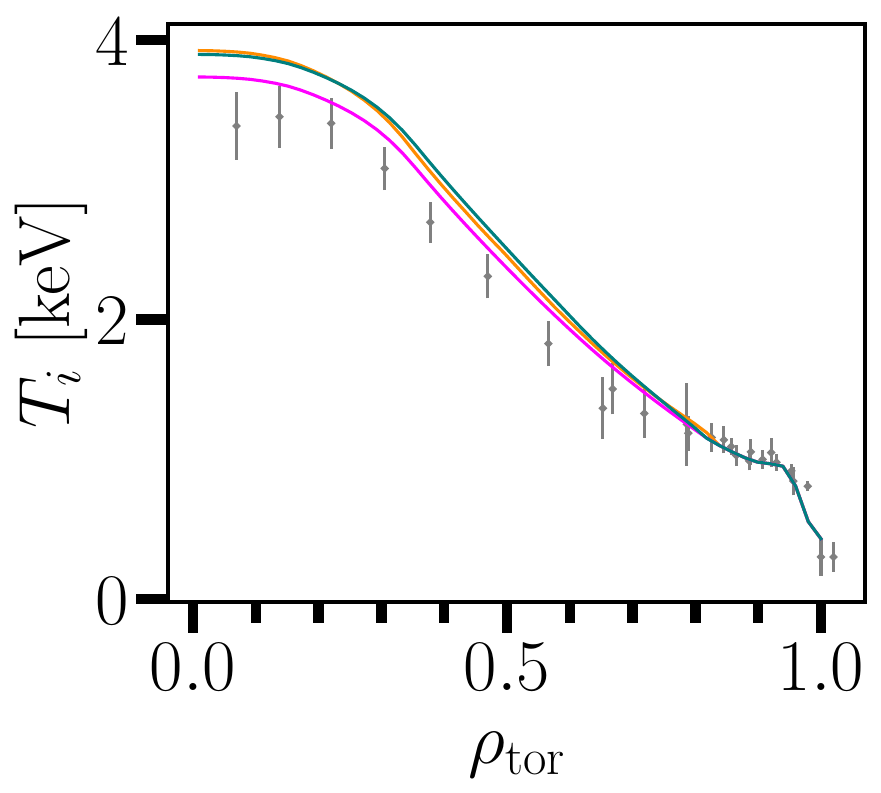}
    }
    \hfill
\caption{Comparison of the steady-state profiles from the simulation of JET\#73342.}
    \label{fig:JET73342Profiles}
\end{figure}

\begin{figure}
    \centering
    \subfigure[]{
        \includegraphics[width=0.45\textwidth]{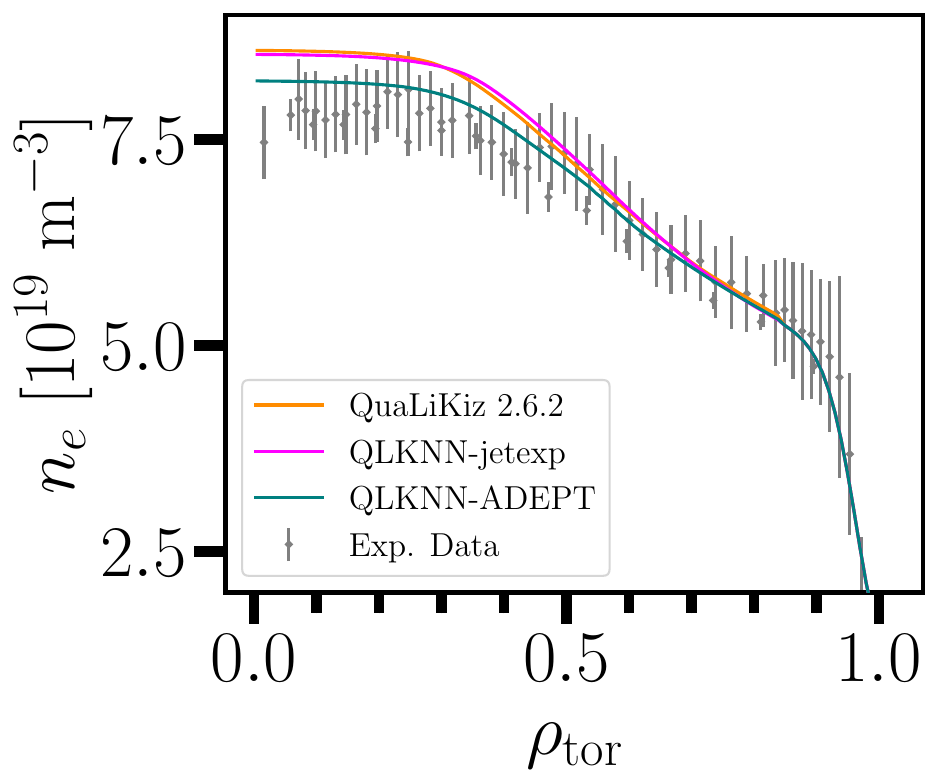}
    }
    \hfill
    \subfigure[]{
        \includegraphics[width=0.45\textwidth]{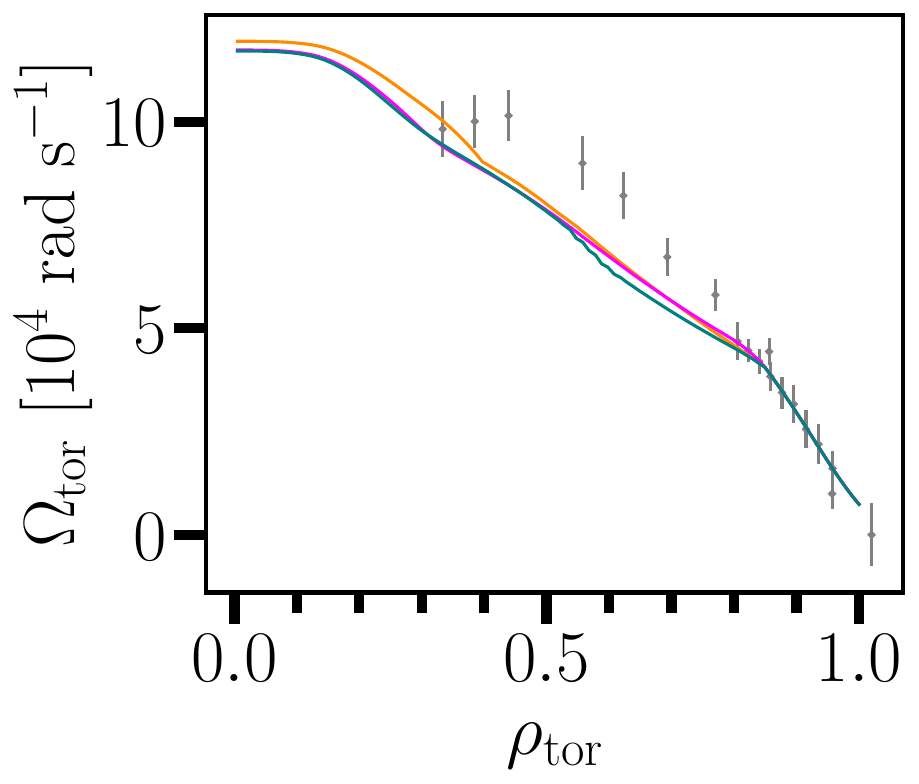}
        
    }
    \vspace{0.5cm} 
    \subfigure[]{
        \includegraphics[width=0.45\textwidth]{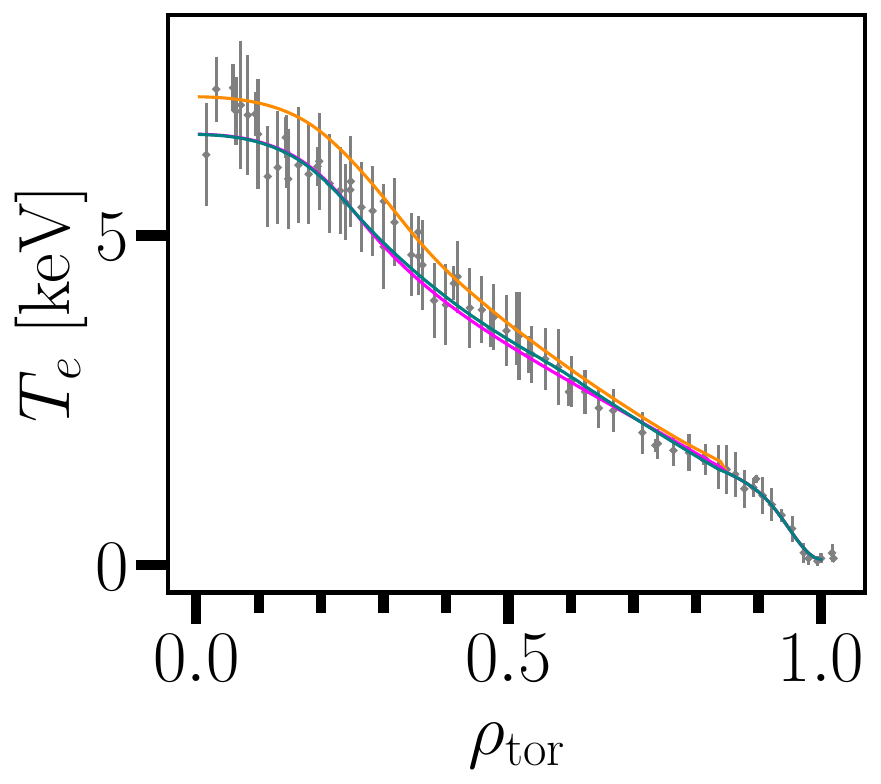}
    }
    \hfill
    \subfigure[]{
        \includegraphics[width=0.45\textwidth]{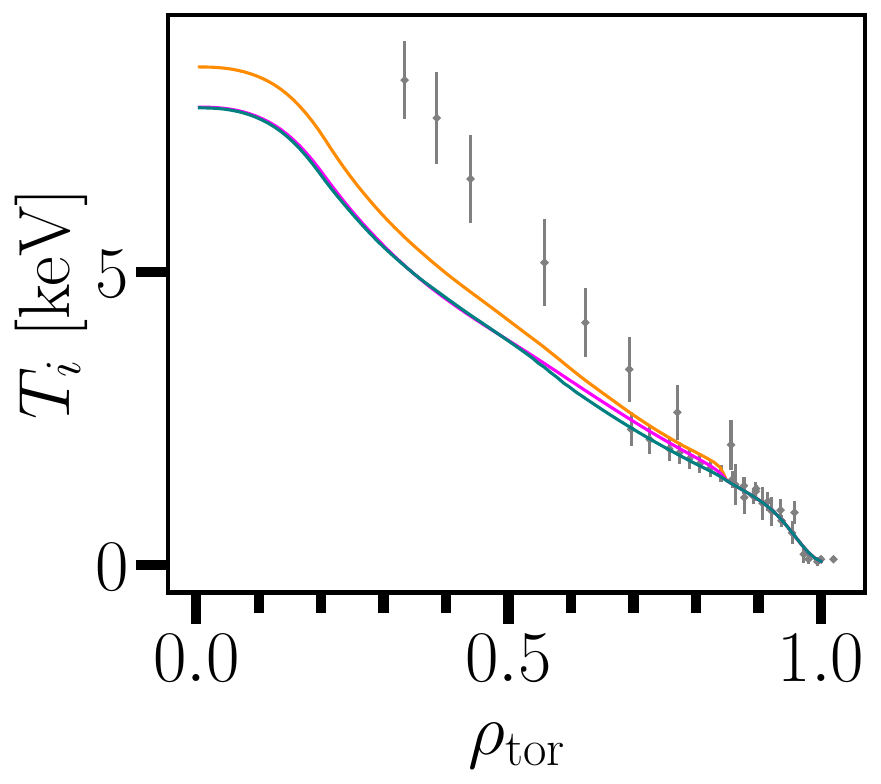}
    }
    \hfill    
\caption{Comparison of the steady-state profiles from the simulation of JET\#92436.}
    \label{fig:JET92436Profiles}
\end{figure}

Figures \ref{fig:JET73342Profiles} and \ref{fig:JET92436Profiles} show the steady state profiles obtained by adopting ADEPT, the original QLKNN-jetexp surrogates of Ref. \cite{jetexp} and the original QuaLiKiz model. The experimental data is shown here for reference, however the purpose of this test is to verify whether the surrogate models are able to reproduce the behaviour of QuaLiKiz within JINTRAC. Table~\ref{tbl:JETTOResultsSummary} provides the profile-averaged relative RMS (RRMS) for these JINTRAC runs using their respective networks, with the QLKNN-jetexp reference given inside the square brackets. The RRMS is computed as:
\begin{equation}
    RRMS = \sqrt{\frac{1}{N} \sum_i \frac{\bigl (Y_{NN,i} -Y_{QLK,i}\bigr)^2}{Y^2_{QLK,i}}}
\end{equation}
where the sum is over the number of radial points, and $Y_{NN,i}$ and $Y_{QLK,i}$ indicate the profiles computed using the NN prediction and QLK respectively.

\begin{table}[tbp]
	\centering
	\caption{Summary table of the JINTRAC predicted profile RRMS within the QuaLiKiz evaluation region. The values are given for the QLKNN-ADEPT simulation, with the reference QLKNN-jetexp simulation provided within the square brackets.}
	\begin{tabular}{l|ccc}
		\toprule[1.5pt]
		& JET\#73342 & JET\#92436  \\
		\midrule
		$T_e$ RRMS & 1.1\% [4.1\%] & 8.0\% [8.8\%] \\
		$T_i$ RRMS & 1.7\% [3.3\%] & 9.1\% [7.3\%] \\
		$n_e$ RRMS & 7.7\% [7.7\%] & 2.7\% [0.7\%] \\
		$\Omega_{\text{tor}}$ RRMS & -- [--] & 4.0\% [3.0\%] \\
		\bottomrule
	\end{tabular}
	\label{tbl:JETTOResultsSummary}
\end{table}

Both the surrogate models considered achieve a match with QuaLiKiz that is better than 10\%. The experiments carried out in this Section suggest that both ADEPT and QLKNN-jetexp models can effectively replace the original transport model as a drop-in replacement for obtaining steady-state profiles, albeit the training dataset acquired by ADEPT was two orders of magnitude smaller than for QLKNN-jetexp.

\section{Summary, conclusions and future work}
\label{sec:Conclusion}

We presented ADEPT (Active Deep Ensembles for Plasma Turbulence), a two-stage physics-informed Active Learning framework for data-efficient surrogate models of gyrokinetic turbulence. ADEPT consists of a classifier NN that learns the boundary manifold between regions that are stable and unstable under linear gyrokinetic turbulence, thus simultaneously providing a model for the critical gradient and limiting the search space of a second surrogate regressing to the turbulent transport fluxes. Using an existing large dataset of QuaLiKiz simulations based on experimental JET discharges \cite{Ho2019,jetexp}, we have demonstrated a sizeable reduction in the size of the training dataset needed to obtain surrogates with integrated performance metrics comparable to surrogates trained by sampling at random. The reduction factor can be up to a factor of 10 or more, and it is due to both the adoption of Active Learning and the physics-informed nature of ADEPT enforced by the classifier, which alone results in increased data efficiency, as only a minority of the QuaLiKiz runs would have been performed deep into the stable regions of the parameter space, thus limiting the need to run costly and uninformative simulations. Compared to previous work, ADEPT delivers similar or superior performance albeit using two order of magnitude fewer data points, and a proportionally lower compute time. We also showed agreement with QuaLiKiz and previous surrogates in parameter scans and in integrated modelling applications to two very different JET discharges. The classifier stage of ADEPT may be relevant for any other model where restricting the parameter space to a certain region with desirable properties is useful, such as the case of building surrogate models of codes modelling magnetohydrodynamic instabilities \cite{Mikhailovskii1998}.

While our results are extremely encouraging, the data volume required to obtain a performing surrogate valid over the sizeable but not extreme parameter space of JET still required of the order of hundreds of thousands simulations, even with Active Learning in a physics-informed setting. Much more efficient strategies should be employed to deliver actionable surrogates of higher fidelity models with high dimensionality and over wide parameter spaces. For instance, acquisition functions that do not employ the surrogate uncertainty should also be considered (e.g., \cite{ash2020deep,sener2018active,holzmüller2022framework}). Moreover, as noted in Section \ref{subsec:TrainingDynamics}, the acquisition function adopted in this work uniquely depends on the sum of the uncertainties of the NNs regressing to the turbulent fluxes. As a result, it is found that the performance of the classifier for a fixed amount of training data does depend on the number of fluxes that contribute to the acquisition function; explicitly accounting for classifier uncertainty in the acquisition function may alleviate or resolve this issue. Lastly, distances between the requested sample and the training distribution may also be explored as a means to quantify uncertainty \cite{malahanobis} as well as to discourage (or encourage) exploration out of the pool distribution in the acquisition function.

\section*{Acknowledgments}
This work has been carried out within the framework of the EUROfusion Consortium, funded by the European Union via the Euratom Research and Training Programme (Grant Agreement No 101052200 EUROfusion). Views and opinions expressed are however those of the author(s) only and do not necessarily reflect those of the European Union or the European Commission. Neither the European Union nor the European Commission can be held responsible for them.

\appendix 
\section{Supervised machine learning with Neural Networks}
\label{app:appendix_NNs}

Neural Networks (NNs) are non-linear, parametric machine learning models based on single units called \emph{neurons}. A collection of neurons is a \emph{layer}, and a collection of layers connected to each other defines the \emph{architecture} of the NN. At each layer, each neuron constructs a learnable linear combination of the outputs from the previous layer using a weight matrix $W$ and a bias $\mathbf{b}$, the parameters of which are indicated collectively as $\mathbf{\theta}$. A non-linear \emph{activation function} $f$ is then applied,
\begin{align}
\label{eq:neuralnet}
\textbf{z}_j &= W_{ij}\textbf{a}_i+\textbf{b}_j  \nonumber\\
\textbf{a}_j &= f(\textbf{z}_j),
\end{align}
where $\textbf{a}_j$, $W_{ij}$ and $\textbf{b}_j$ are the output, weight matrix and bias of the current layer, while $\textbf{a}_i$ is the output of the previous layer. The first layer ingests the data $\textbf{x}$, so that $\textbf{a}_0\equiv\textbf{x}$. If the i-th and j-th layers contain M and N neurons respectively, then $W_{ij}$ will be an MxN matrix. As the layers of a NN may have different number of neurons, subsequent layers are linked by weight matrices $W$ with suitable dimensions.

In this work, we are interested in using NNs for supervised learning. In supervised learning, a machine learning algorithm is trained on a dataset for which both $x_{train}$ and $y_{train}$ are known. In probabilistic terms, the algorithm learns the distribution $p(y|x)$ of labels $y$ given an input $x$. During training, the discrepancy between the output of the NN $\hat{y}|\mathbf{x_{train}}$ and the true output $y_{train}$ is quantified by means of a loss functions, and this information is used to adjust the weights and biases of the NN.

Supervised learning includes both \emph{regression} and in \emph{classification} tasks. For regression, the labels $y\in \mathbb{R}$ are real numbers which can assume any value in the real domain. Instead, in a classification task the labels are discrete.

\section{NN integrated performance metrics}
\label{app:perfmetrics}

We evaluate the surrogates in terms of the R$^2$ score for the regressors and the F1 score for the classifier, defined as follows:
\begin{align}
    \text{R}^2 & = 1- \frac{\sum_i (y_i-\hat{y}_i)^2}{\sum_i(y_i-\overline{y})^2}\\
    \text{F1} & =  2 \ \text{x}  \ \frac{\text{Precision} \ \text{x}\ \text{Recall}}{\text{Precision} + \text{Recall}}\\
    \text{Precision} &= \frac{TP}{TP+FP}\\
    \text{Recall}& = \frac{TP}{TP+FN}    
\end{align}
where $\overline{y}$ is the mean of the target flux in the test dataset, and TP, FP and FN are the true positives, false positive and false negatives. For all metrics, higher values indicate better quality of the surrogates, with a maximum value of 1.

The F1 score is more suitable than Accuracy to evaluate performance on imbalanced datasets like ours, where only 25\% of inputs are unstable, and it is in general recommended for AL workflows as the relative proportion of positive and negative labels (i.e., unstable and stable regions in our case) is unknown a priori. To be more precise, Accuracy provides a measure of correct predictions across both positive and negative labels. In situations where datasets are heavily imbalanced, the model might focus solely on performing well with the majority class, resulting in seemingly good or even almost perfect performance, while the minority class is never predicted correctly. Thus, Accuracy alone does not shed light on the occurrences of false positives and false negatives, which are equally important to capture. Recall specifically addresses false negatives, indicating how often the model erroneously classifies something as negative when it's actually positive. Conversely, precision deals with false positives, revealing how frequently the model incorrectly labels something as positive when it's truly negative. Therefore, in regions of instability, having a high recall is beneficial as it maximizes the detection of truly unstable points while minimizing false negatives. Conversely, in stable regions a high precision is valuable because it reduces the occurrence of spurious fluxes (see also Appendix G of \cite{plassche2020}). The F1 score is the geometric average of Precision and Recall, thus capturing an overall performance across both regions while accounting for imbalanced data.

\section{Further parameter scan validation}
\label{app:OtherScans}

\begin{figure}
    \centering
    \includegraphics[width=0.9\textwidth]{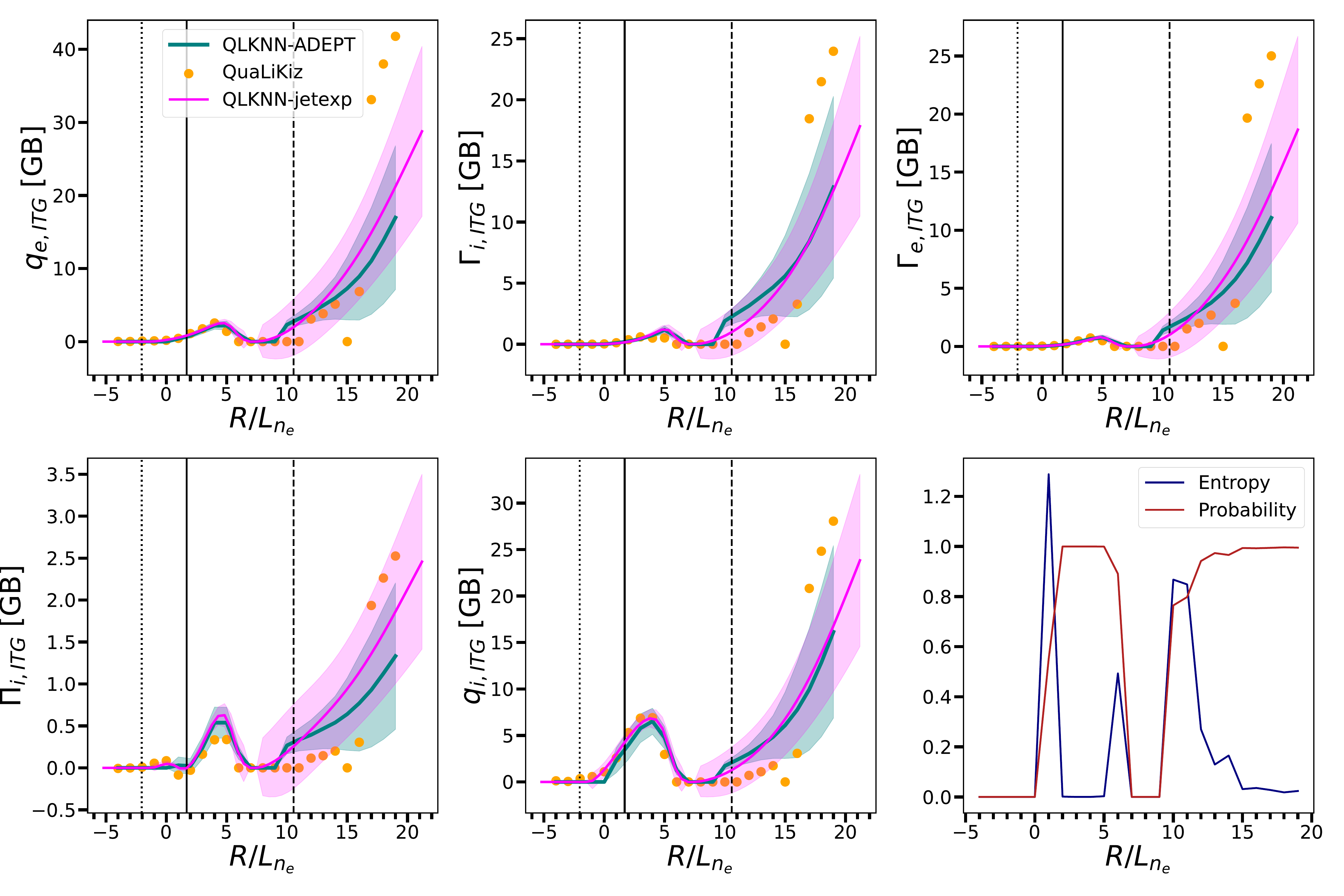}
    \caption{Parameter scans in $R/L_{n_e}$ for the five ITG fluxes used in this work. }
    \label{fig:otherparamscans_ane}
\end{figure}

\begin{figure}
    \centering
    \includegraphics[width=0.9\textwidth]{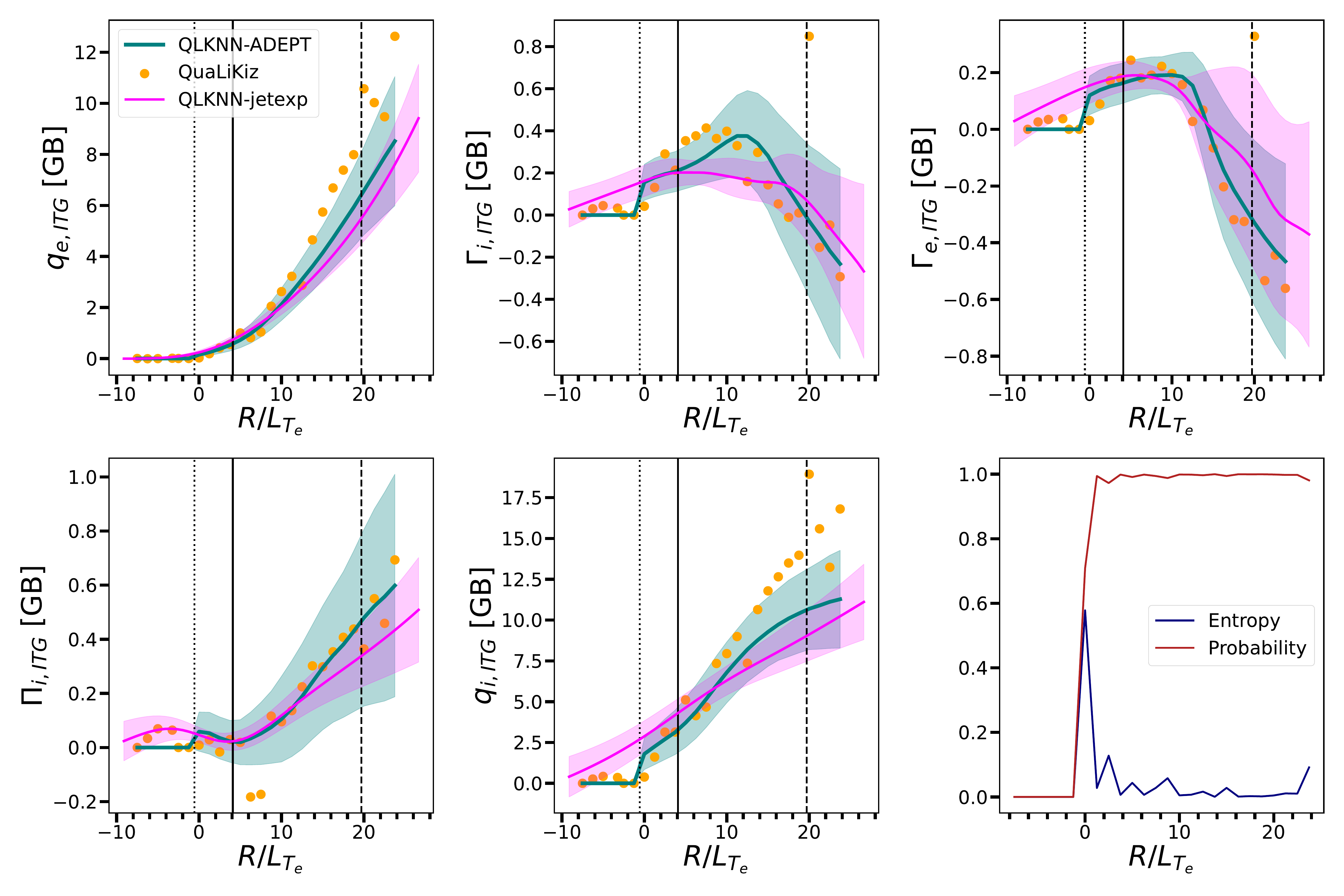}
    \caption{Parameter scans in $R/L_{T_e}$ for the five ITG fluxes used in this work.}
    \label{fig:otherparamscans_ate}
\end{figure}

\begin{figure}
    \centering
    \includegraphics[width=0.9\textwidth]{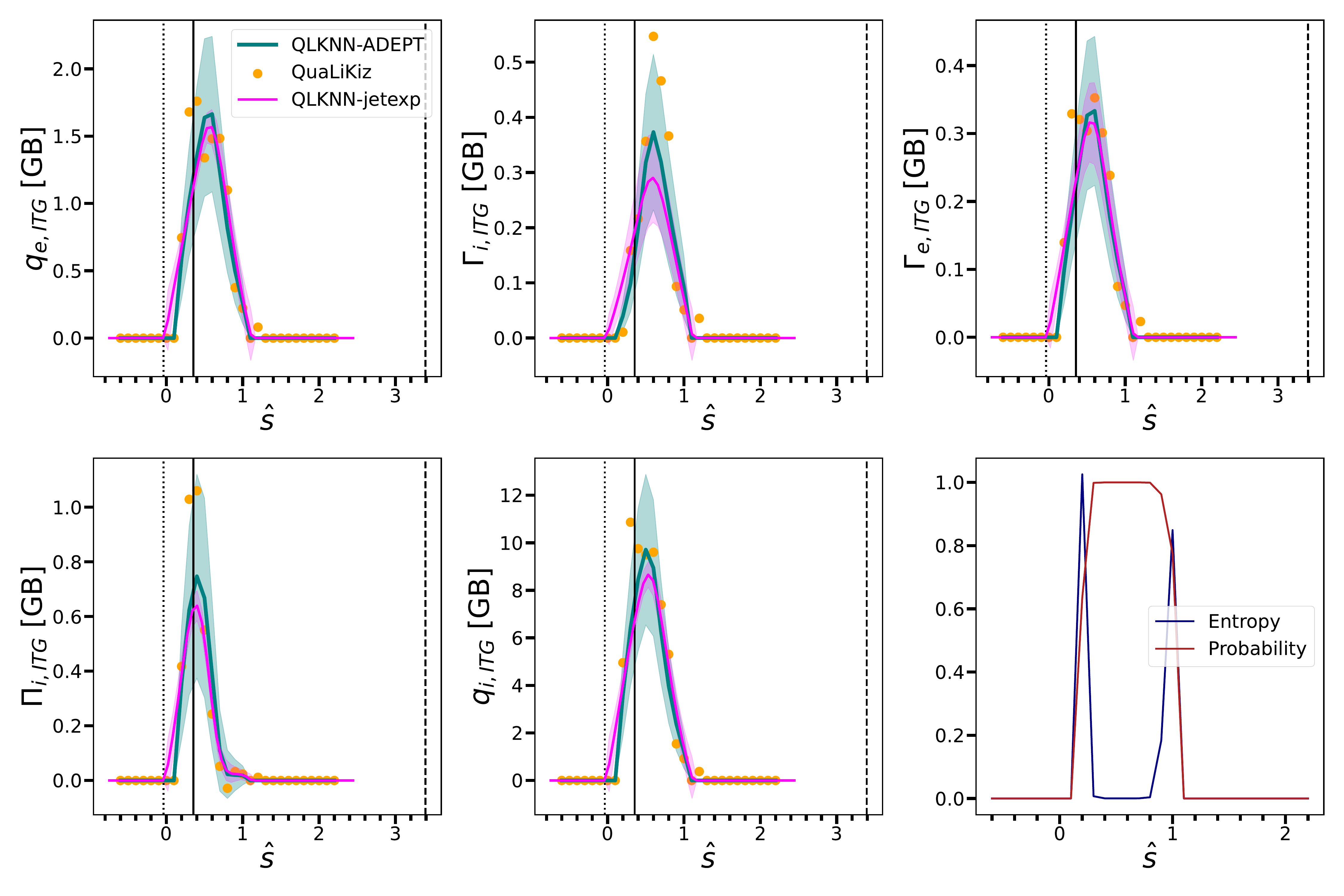}
    \caption{Parameter scans in $\hat{s}$ for the five ITG fluxes used in this work.}
    \label{fig:otherparamscans_smag}
\end{figure}
\begin{figure}
    \centering
    \includegraphics[width=0.9\textwidth]{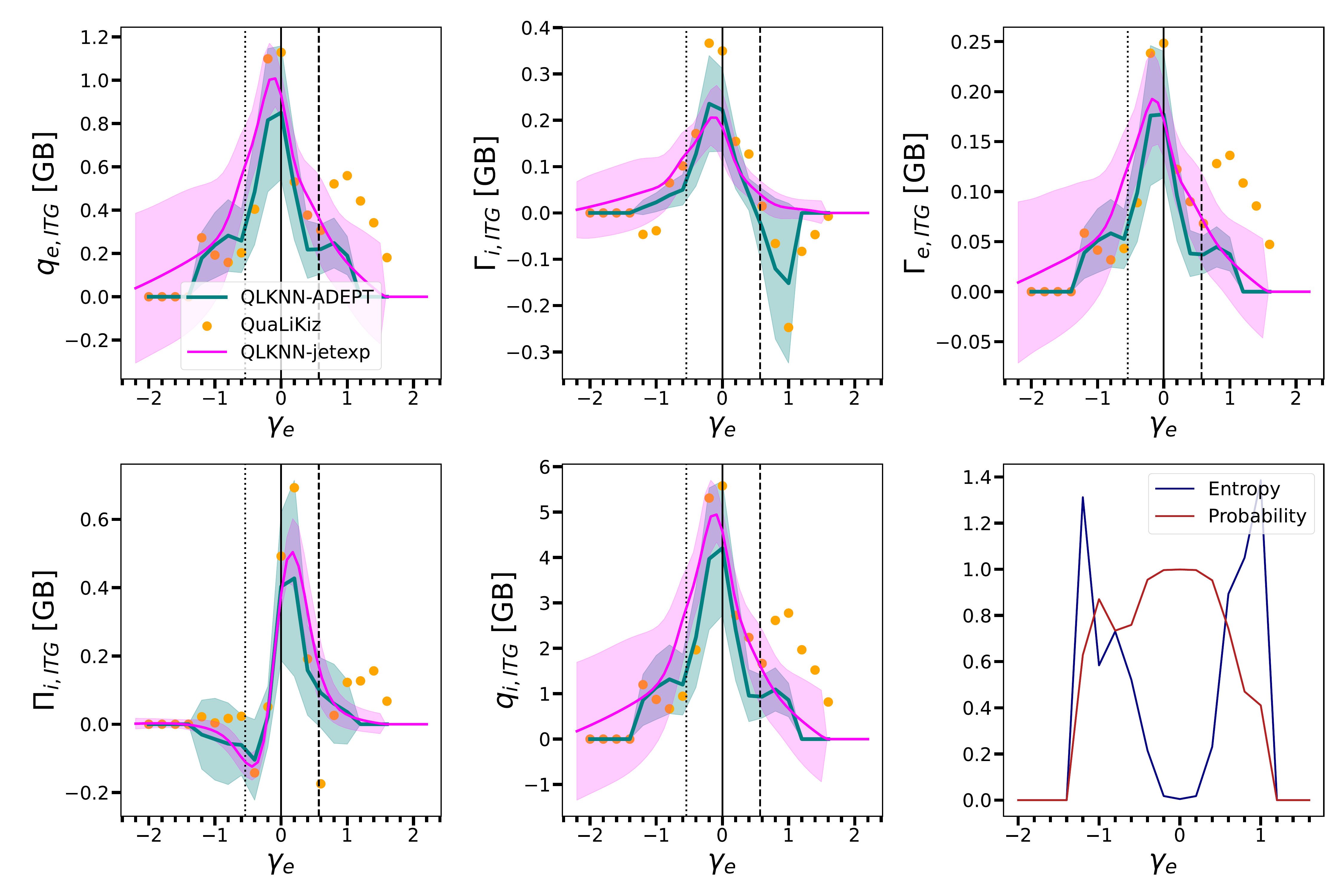}
    \caption{Parameter scans in $\gamma_E$ for the five ITG fluxes used in this work. }
    \label{fig:otherparamscans_gammae}
\end{figure}

Further parameter scans performed using ADEPT are shown in Figures \ref{fig:otherparamscans_ane}, \ref{fig:otherparamscans_ate}, \ref{fig:otherparamscans_smag}, \ref{fig:otherparamscans_gammae}. In all the Figures, the QuaLiKiz runs are shown in green, while the predictions of the surrogates are shown in red and the dashed areas indicate the 1$\sigma$ confidence levels. Dotted, solid and dashed lines indicate the 2.5\%, 50\% and 97.5\% of the distribution of the parameter being scanned. The bottom right panel shows the uncertainty for the Deep Ensemble classifier in terms of probability of an input being unstable and entropy of the ensemble. 

\newpage
\section*{References}
\bibliographystyle{unsrt}
\bibliography{refs}

\begin{thebibliography}{10}

\bibitem{aTurbulentTransport-Callen}
J.~D. Callen.
\newblock Transport processes in magnetically confined plasmas.
\newblock {\em Physics of Fluids B: Plasma Physics}, 4(7):2142--2154, 1992.

\bibitem{aToroidalConfinement-Hinton}
F.~L. Hinton and R.~D. Hazeltine.
\newblock Theory of plasma transport in toroidal confinement systems.
\newblock {\em Rev. Mod. Phys.}, 48:239--308, 04 1976.

\bibitem{aJINTRAC-Romanelli}
M~Romanelli, G~Corrigan, V~Parail, Sven Wiesen, Roberto Ambrosino, P~Da~Silva Aresta~Belo, Luca Garzotti, P~Harting, F~K{\"o}chl, Tuomas Koskela, L~Lauro-Taroni, Chiara Marchetto, Massimiliano Mattei, E~Militello-Asp, M~Nave, Stanislas Pamela, A~Salmi, P~Strand, and G~Szepesi.
\newblock {JINTRAC}: {A} system of codes for integrated simulation of tokamak scenarios.
\newblock {\em Plasma and Fusion Research}, 9, 01 2014.

\bibitem{Felici2011Raptor}
F.~{Felici}, O.~{Sauter}, S.~{Coda}, B.~P. {Duval}, T.~P. {Goodman}, J.~M. {Moret}, J.~I. {Paley}, and {TCV Team}.
\newblock {Real-time physics-model-based simulation of the current density profile in tokamak plasmas}.
\newblock {\em Nuclear Fusion}, 51(8):083052, August 2011.

\bibitem{Bourdelle2015}
C~Bourdelle, J~Citrin, B~Baiocchi, A~Casati, P~Cottier, X~Garbet, and F~Imbeaux and.
\newblock Core turbulent transport in tokamak plasmas: bridging theory and experiment with {QuaLiKiz}.
\newblock {\em Plasma Physics and Controlled Fusion}, 58(1):014036, December 2015.

\bibitem{Citrin2017}
J~Citrin, C~Bourdelle, F~J Casson, C~Angioni, N~Bonanomi, Y~Camenen, X~Garbet, L~Garzotti, T~G\"{o}rler, O~G\"{u}rcan, F~Koechl, F~Imbeaux, O~Linder, K~van~de Plassche, P~Strand, and G~Szepesi and.
\newblock Tractable flux-driven temperature, density, and rotation profile evolution with the quasilinear gyrokinetic transport model {QuaLiKiz}.
\newblock {\em Plasma Physics and Controlled Fusion}, 59(12):124005, November 2017.

\bibitem{plassche2020}
K.~L. van~de Plassche, J.~Citrin, C.~Bourdelle, Y.~Camenen, F.~J. Casson, V.~I. Dagnelie, F.~Felici, A.~Ho, S.~Van Mulders, JET Contributors, and et~al.
\newblock Fast modeling of turbulent transport in fusion plasmas using neural networks.
\newblock {\em AIP Publishing}, Feb 2020.

\bibitem{Staebler2007}
G.~M. Staebler, J.~E. Kinsey, and R.~E. Waltz.
\newblock A theory-based transport model with comprehensive physics.
\newblock {\em Physics of Plasmas}, 14(5):055909, May 2007.

\bibitem{Staebler2010}
G.~M. Staebler and J.~E. Kinsey.
\newblock Electron collisions in the trapped gyro-landau fluid transport model.
\newblock {\em Physics of Plasmas}, 17(12):122309, December 2010.

\bibitem{Meneghini2017}
O.~Meneghini, S.P. Smith, P.B. Snyder, G.M. Staebler, J.~Candy, E.~Belli, L.~Lao, M.~Kostuk, T.~Luce, T.~Luda, J.M. Park, and F.~Poli.
\newblock Self-consistent core-pedestal transport simulations with neural network accelerated models.
\newblock {\em Nuclear Fusion}, 57(8):086034, July 2017.

\bibitem{Felici2018}
F.~Felici, J.~Citrin, A.A. Teplukhina, J.~Redondo, C.~Bourdelle, F.~Imbeaux, O.~Sauter, and and.
\newblock Real-time-capable prediction of temperature and density profiles in a tokamak using {RAPTOR} and a first-principle-based transport model.
\newblock {\em Nuclear Fusion}, 58(9):096006, July 2018.

\bibitem{VanMulders2021}
S.~Van Mulders, F.~Felici, O.~Sauter, J.~Citrin, A.~Ho, M.~Marin, and K.L. van~de Plassche.
\newblock Rapid optimization of stationary tokamak plasmas in {RAPTOR}: demonstration for the {ITER} hybrid scenario with neural network surrogate transport model {QLKNN}.
\newblock {\em Nuclear Fusion}, 61(8):086019, July 2021.

\bibitem{Meneghini2020}
O.~Meneghini, G.~Snoep, B.C. Lyons, J.~McClenaghan, C.S. Imai, B.~Grierson, S.P. Smith, G.M. Staebler, P.B. Snyder, J.~Candy, E.~Belli, L.~Lao, J.M. Park, J.~Citrin, T.L. Cordemiglia, A.~Tema, and S.~Mordijck.
\newblock Neural-network accelerated coupled core-pedestal simulations with self-consistent transport of impurities and compatible with {ITER} {IMAS}.
\newblock {\em Nuclear Fusion}, 61(2):026006, December 2020.

\bibitem{RodriguezFernandez2022}
P.~Rodriguez-Fernandez, N.T. Howard, and J.~Candy.
\newblock Nonlinear gyrokinetic predictions of {SPARC} burning plasma profiles enabled by surrogate modeling.
\newblock {\em Nuclear Fusion}, 62(7):076036, May 2022.

\bibitem{Farca2022}
Ionu{\c{t}}-Gabriel Farca{\c{s}}, Gabriele Merlo, and Frank Jenko.
\newblock A general framework for quantifying uncertainty at scale.
\newblock {\em Communications Engineering}, 1(1), December 2022.

\bibitem{jetexp}
A.~Ho, J.~Citrin, C.~Bourdelle, Y.~Camenen, F.~J. Casson, K.~L. van~de Plassche, H.~Weisen, and JET Contributors.
\newblock Neural network surrogate of qualikiz using jet experimental data to populate training space.
\newblock {\em AIP Publishing}, Mar 2021.

\bibitem{Narita2021}
E.~Narita, M.~Honda, M.~Nakata, M.~Yoshida, and N.~Hayashi.
\newblock Quasilinear turbulent particle and heat transport modelling with a neural-network-based approach founded on gyrokinetic calculations and experimental data.
\newblock {\em Nuclear Fusion}, 61(11):116041, October 2021.

\bibitem{aGKW-Peeters2009}
A.G. Peeters, Y.~Camenen, F.J. Casson, W.A. Hornsby, A.P. Snodin, D.~Strintzi, and G.~Szepesi.
\newblock The nonlinear gyro-kinetic flux tube code gkw.
\newblock {\em Computer Physics Communications}, 180(12):2650--2672, 2009.

\bibitem{Citrin_2023}
J.~Citrin, P.~Trochim, T.~Goerler, D.~Pfau, K.~L. van~de Plassche, and F.~Jenko.
\newblock Fast transport simulations with higher-fidelity surrogate models for {ITER}.
\newblock {\em Physics of Plasmas}, 30(6), jun 2023.

\bibitem{aTwoStep-Kremers}
Bart J.~J. Kremers, Jonathan Citrin, Aaron Ho, and Karel~L. van~der Plassche.
\newblock Two-step clustering for data reduction combining dbscan and k-means clustering.
\newblock {\em Contributions to Plasma Physics}, 63(5-6):e202200177, 2023.

\bibitem{barr2022}
J.~Barr, T.~Madula, L.~Zanisi, A.~Ho, J.~Citrin, V.~Gopakumar, and {JET Contributors}.
\newblock An active learning pipeline for surrogate models of gyrokinetic turbulence.
\newblock In {\em 48th EPS Conference on Plasma Physics 27 June - 1 July 2022}, Europhysics conference abstracts. European Physical Society (EPS), 2022.

\bibitem{hornsby2023gaussian}
William Hornsby, Ander Gray, James Buchanan, Bhavin Patel, Daniel Kennedy, Francis Casson, Colin Roach, Mikkel Lykkegaard, Huy Nguyen, Nikolaos Papadimas, Ben Fourcin, and Jordan Hart.
\newblock Gaussian process regression models for the properties of micro-tearing modes in spherical tokamak, 2023.

\bibitem{Aggarwal_survey}
{Charu C.} Aggarwal, Xiangnan Kong, Quanquan Gu, Jiawei Han, and {Philip S.} Yu.
\newblock {\em Active learning: A survey}, pages 571--605.
\newblock CRC Press, January 2014.
\newblock Publisher Copyright: {\textcopyright} 2015 by Taylor and Francis Group, LLC.

\bibitem{aBO-Jarvinen2022}
A.E. J{\"a}rvinen, T.~F{\"u}l{\"o}p, E.~Hirvijoki, M.~Hoppe, A.~Kit, and J.~{\AA}str{\"o}m.
\newblock Bayesian approach for validation of runaway electron simulations.
\newblock {\em Journal of Plasma Physics}, 88(6):905880612, 2022.

\bibitem{aBayesianPlasmaBoundary-Skvara}
V{\'{i}}t {\v{S}}kv{\'{a}}ra, V{\'{a}}clav {\v{S}}m{\'{i}}dl, and Jakub Urban.
\newblock Robust sparse linear regression for tokamak plasma boundary estimation using variational {Bayes}.
\newblock {\em Journal of Physics: Conference Series}, 1047:012015, 06 2018.

\bibitem{chung2020offline}
Youngseog Chung, Ian Char, Willie Neiswanger, Kirthevasan Kandasamy, Andrew~Oakleigh Nelson, Mark~D Boyer, Egemen Kolemen, and Jeff Schneider.
\newblock Offline contextual bayesian optimization for nuclear fusion, 2020.

\bibitem{MacKay1992}
David J.~C. MacKay.
\newblock {Information-Based Objective Functions for Active Data Selection}.
\newblock {\em Neural Computation}, 4(4):590--604, 07 1992.

\bibitem{lakshminarayanan2017simple}
Balaji Lakshminarayanan, Alexander Pritzel, and Charles Blundell.
\newblock Simple and scalable predictive uncertainty estimation using deep ensembles, 2017.

\bibitem{Ho2019}
A.~Ho, J.~Citrin, F.~Auriemma, C.~Bourdelle, F.J. Casson, Hyun-Tae Kim, P.~Manas, G.~Szepesi, and H.~Weisen and.
\newblock Application of gaussian process regression to plasma turbulent transport model validation via integrated modelling.
\newblock {\em Nuclear Fusion}, 59(5):056007, March 2019.

\bibitem{Rasmussen2004}
Carl~Edward Rasmussen.
\newblock Gaussian processes in machine learning.
\newblock In {\em Advanced Lectures on Machine Learning}, pages 63--71. Springer Berlin Heidelberg, 2004.

\bibitem{Galaxy_zoo_AL}
Mike {Walmsley}, Lewis {Smith}, Chris {Lintott}, Yarin {Gal}, Steven {Bamford}, Hugh {Dickinson}, Lucy {Fortson}, Sandor {Kruk}, Karen {Masters}, Claudia {Scarlata}, Brooke {Simmons}, Rebecca {Smethurst}, and Darryl {Wright}.
\newblock {Galaxy Zoo: probabilistic morphology through Bayesian CNNs and active learning}.
\newblock {\em Monthly Notices of the Royal Astronomical Society}, 491(2):1554--1574, January 2020.

\bibitem{McKay1979}
M.~D. McKay, R.~J. Beckman, and W.~J. Conover.
\newblock A comparison of three methods for selecting values of input variables in the analysis of output from a computer code.
\newblock {\em Technometrics}, 21(2):239, May 1979.

\bibitem{Maxentropy_NP_Hard}
Chun-Wa Ko, Jon Lee, and Maurice Queyranne.
\newblock {An Exact Algorithm for Maximum Entropy Sampling}.
\newblock {\em Operations Research}, 43(4):684--691, August 1995.

\bibitem{holzmüller2022framework}
David Holzmüller, Viktor Zaverkin, Johannes Kästner, and Ingo Steinwart.
\newblock A framework and benchmark for deep batch active learning for regression.
\newblock {\em arXiv}, 2022.

\bibitem{Ren_BMDAL_review}
Pengzhen Ren, Yun Xiao, Xiaojun Chang, Po-Yao Huang, Zhihui Li, Brij~B. Gupta, Xiaojiang Chen, and Xin Wang.
\newblock A survey of deep active learning.
\newblock {\em ACM Comput. Surv.}, 54(9), oct 2021.

\bibitem{Soleyimani21}
Ava~P. Soleimany, Alexander Amini, Samuel Goldman, Daniela Rus, Sangeeta~N. Bhatia, and Connor~W. Coley.
\newblock Evidential deep learning for guided molecular property prediction and discovery.
\newblock {\em ACS Central Science}, 7(8):1356--1367, 2021.

\bibitem{ALBias}
Sanjoy Dasgupta and Daniel Hsu.
\newblock Hierarchical sampling for active learning.
\newblock In {\em Proceedings of the 25th International Conference on Machine Learning}, ICML '08, page 208–215, New York, NY, USA, 2008. Association for Computing Machinery.

\bibitem{nguyen2015deep}
Anh Nguyen, Jason Yosinski, and Jeff Clune.
\newblock Deep neural networks are easily fooled: High confidence predictions for unrecognizable images, 2015.

\bibitem{gawlikowski2022survey}
Jakob Gawlikowski, Cedrique Rovile~Njieutcheu Tassi, Mohsin Ali, Jongseok Lee, Matthias Humt, Jianxiang Feng, Anna Kruspe, Rudolph Triebel, Peter Jung, Ribana Roscher, Muhammad Shahzad, Wen Yang, Richard Bamler, and Xiao~Xiang Zhu.
\newblock A survey of uncertainty in deep neural networks, 2022.

\bibitem{pmlr-v70-guo17a}
Chuan Guo, Geoff Pleiss, Yu~Sun, and Kilian~Q. Weinberger.
\newblock On calibration of modern neural networks.
\newblock In Doina Precup and Yee~Whye Teh, editors, {\em Proceedings of the 34th International Conference on Machine Learning}, volume~70 of {\em Proceedings of Machine Learning Research}, pages 1321--1330. PMLR, 06--11 Aug 2017.

\bibitem{angelopoulos2022gentle}
Anastasios~N. Angelopoulos and Stephen Bates.
\newblock A gentle introduction to conformal prediction and distribution-free uncertainty quantification, 2022.

\bibitem{proper_scoring_rule}
Tilmann Gneiting and Adrian~E Raftery.
\newblock Strictly proper scoring rules, prediction, and estimation.
\newblock {\em Journal of the American Statistical Association}, 102(477):359--378, 2007.

\bibitem{Brcker2007}
Jochen Br\"{o}cker and Leonard~A. Smith.
\newblock Scoring probabilistic forecasts: The importance of being proper.
\newblock {\em Weather and Forecasting}, 22(2):382--388, April 2007.

\bibitem{gustafsson2020evaluating}
Fredrik~K. Gustafsson, Martin Danelljan, and Thomas~B. Schön.
\newblock Evaluating scalable bayesian deep learning methods for robust computer vision, 2020.

\bibitem{Yudin_2023}
Yehor Yudin, David Coster, Udo von Toussaint, and Frank Jenko.
\newblock {Epistemic and Aleatoric Uncertainty Quantification and Surrogate Modelling in High-Performance Multiscale Plasma Physics Simulations}.
\newblock In {\em ICCS 2023}, pages 572--586. Springer Nature Switzerland, 2023.

\bibitem{shannonMathematicalTheoryCommunication1948}
Claude~Elwood Shannon.
\newblock A {{Mathematical Theory}} of {{Communication}}.
\newblock {\em Bell System Technical Journal}, 27(3):379--423, 1948.

\bibitem{JenkoGENE}
F.~{Jenko}, W.~{Dorland}, M.~{Kotschenreuther}, and B.~N. {Rogers}.
\newblock {Electron temperature gradient driven turbulence}.
\newblock {\em Physics of Plasmas}, 7(5):1904--1910, May 2000.

\bibitem{aITER-JINTRAC-Asp2022}
E.~Militello Asp, G.~Corrigan, P.~da~Silva Aresta~Belo, L.~Garzotti, D.M. Harting, F.~Köchl, V.~Parail, M.~Cavinato, A.~Loarte, M.~Romanelli, and R.~Sartori.
\newblock {JINTRAC} integrated simulations of {ITER} scenarios including fuelling and divertor power flux control for {H}, {He} and {DT} plasmas.
\newblock {\em Nuclear Fusion}, 62(12):126033, oct 2022.

\bibitem{Marin2020}
M.~Marin, J.~Citrin, C.~Bourdelle, Y.~Camenen, F.J. Casson, A.~Ho, F.~Koechl, and M.~Maslov and.
\newblock First-principles-based multiple-isotope particle transport modelling at {JET}.
\newblock {\em Nuclear Fusion}, 60(4):046007, February 2020.

\bibitem{Mikhailovskii1998}
A~B Mikhailovskii.
\newblock Generalized {MHD} for numerical stability analysis of high-performance plasmas in tokamaks.
\newblock {\em Plasma Physics and Controlled Fusion}, 40(11):1907--1920, November 1998.

\bibitem{ash2020deep}
Jordan~T. Ash, Chicheng Zhang, Akshay Krishnamurthy, John Langford, and Alekh Agarwal.
\newblock Deep batch active learning by diverse, uncertain gradient lower bounds, 2020.

\bibitem{sener2018active}
Ozan Sener and Silvio Savarese.
\newblock Active learning for convolutional neural networks: A core-set approach, 2018.

\bibitem{malahanobis}
Prasanta~Chandra Malahanobis.
\newblock Reprint of: Mahalanobis, p.c. (1936) "on the generalised distance in statistics.".
\newblock {\em Sankhya A}, 80(S1):1--7, December 2018.

\end{thebibliography}

\end{document}